\pdfoutput=1
\documentclass[12pt,a4paper,article]{memoir} 

\usepackage{tgpagella}

\usepackage[utf8]{inputenc}
\usepackage{amsmath,amssymb,mathtools,amsthm}
\usepackage[margin=1in]{geometry}
\usepackage{dsfont, lmodern}
\usepackage{braket}
\usepackage{booktabs}
\usepackage{enumerate}
\usepackage{marginnote}
\usepackage{MnSymbol}
\usepackage{stmaryrd}
\usepackage{longtable}

\usepackage[most]{tcolorbox}
\usepackage{algorithm}
\usepackage{algorithmic}
\usepackage[normalem]{ulem}
\usepackage{cancel,xcolor}
\usepackage{orcidlink}
\usepackage{enumitem}
\usepackage{float}
\usepackage{listings}
\usepackage{multirow}
\usepackage{subcaption}
\usepackage{placeins}

\definecolor{jlkeyword}{RGB}{33,74,135}
\definecolor{jlcomment}{RGB}{120,120,120}
\definecolor{jlidentifier}{RGB}{0,102,0}
\definecolor{jlmacro}{RGB}{160,32,240}
\definecolor{jlstring}{RGB}{196,26,22}

\lstdefinelanguage{JuliaCustom}{
  sensitive=true,
  morekeywords={abstract,baremodule,begin,break,catch,const,continue,do,else,elseif,end,export,finally,for,function,if,import,importall,let,local,macro,module,mutable,primitive,quote,return,struct,try,using,while},
  morecomment=[l]{\#},
  morecomment=[n]{\#=}{=\#},
  morestring=[b]" , 
  morestring=[b]'   
}

\lstdefinestyle{code}{
  columns=fullflexible,
  keepspaces=true,
  language=JuliaCustom,
  alsoletter={@},
  morekeywords=[2]{@ncmonoid,@time,@show,@assert, @pcmonoid, @comms}, 
  keywordstyle=\color{jlkeyword}\bfseries,
  keywordstyle=[2]\color{jlmacro}\bfseries,
  commentstyle=\color{jlcomment}\itshape,
  identifierstyle=\color{jlidentifier},
  stringstyle=\color{jlstring},
  showstringspaces=false,
  xleftmargin=1.5cm,
  basicstyle=\ttfamily\small,
  frame=single,
  framesep=4pt,
  rulecolor=\color{black!15},
  breaklines=true,
  tabsize=4,
  mathescape=true,
  literate={ρ}{{$\rho$}}1
           {σ}{{$\sigma$}}1
           {α}{{$\alpha$}}1
           {β}{{$\beta$}}1
           {γ}{{$\gamma$}}1
           {δ}{{$\delta$}}1
           {ϵ}{{$\epsilon$}}1
           {λ}{{$\lambda$}}1
           {μ}{{$\mu$}}1
           {π}{{$\pi$}}1
           {Π}{{$\Pi$}}1
           {φ}{{$\phi$}}1
           {θ}{{$\theta$}}1
           {ω}{{$\omega$}}1
           {≤}{{$\le$}}1
           {≥}{{$\ge$}}1
           {≠}{{$\ne$}}1
           {∈}{{$\in$}}1
           {∀}{{$\forall$}}1
           {∨}{{$\vee$}}1
           {∧}{{$\wedge$}}1
           {¬}{{$\neg$}}1
}
\lstnewenvironment{code}{\lstset{style=code}}{} 

\newtcolorbox{juliacode}[1][]{
  enhanced,
  breakable,
  colback=gray!5,
  colframe=gray!40,
  boxrule=0.5pt,
  arc=6pt,
  left=8pt,
  right=8pt,
  top=6pt,
  bottom=6pt,
  fonttitle=\bfseries,
  title=#1
}

\newtheorem{theorem}{Theorem}

\newtheorem{example}[theorem]{Example}

\newcommand{\one}{\mathbf{1}}

\newcommand{\innerprod}[2]{\langle #1, #2 \rangle}

\newcommand{\st}{\operatorname{s.t.}}

\newcommand{\monoid}{\mathds M}

\newcommand{\mult}{ }

\colorlet{summarycolor}{yellow!70!red!90!blue!80!white}
\colorlet{algorithmcolor}{blue!50!white}
\colorlet{softwarecolor}{red!50!white}

\newtcbtheorem[auto counter,number within=section]{summary}%
  {Highlight}{fonttitle=\bfseries\upshape, fontupper=\slshape,
     arc=0mm, colback=blue!5!white,colframe=summarycolor}{theorem}

\newcommand\blueout{\bgroup\markoverwith
{\textcolor{blue}{\rule[.5ex]{2pt}{0.4pt}}}\ULon}
\newcommand{\mathsout}[1]
{\bgroup\mathchoice
  {\sbox0{$\displaystyle{#1}$}%
    \usebox0\hspace{-\wd0}%
    \rule[0.5\ht0-0.5\dp0-.5pt]{\wd0}{1pt}}%
  {\sbox0{$\textstyle{#1}$}%
    \usebox0\hspace{-\wd0}%
    \rule[0.5\ht0-0.5\dp0-.5pt]{\wd0}{1pt}}%
  {\sbox0{$\scriptstyle{#1}$}%
    \usebox0\hspace{-\wd0}%
    \rule[0.5\ht0-0.5\dp0-.5pt]{\wd0}{1pt}}%
  {\sbox0{$\scriptscriptstyle{#1}$}%
    \usebox0\hspace{-\wd0}%
    \rule[0.5\ht0-0.5\dp0-.5pt]{\wd0}{1pt}}%
\egroup}
\let \vec \textbf
\newcommand{\id}{\mathbf{1}}

\newcommand{\nn}{\nonumber}
\newcommand{\tr}{\operatorname{tr}}

\let\vec\mathbf
\let\epsilon\varepsilon

\title{\Large PCPOP.jl: A Julia package for partially commutative polynomial optimization}

\usepackage{authblk} 

\author[1]{Moisés Bermejo Morán}
\author[2]{Abhishek Mishra}

\affil[1]{\large Department of Computer Science, School of Computing and Data Science, The University of Hong Kong, Hong Kong}
\affil[2]{\large Laboratoire d'Information Quantique, Université libre de Bruxelles, Belgium}

\date{\vspace{-5ex}} 

\begin{document}

\maketitle

\begin{abstract}

Here we present \texttt{PCPOP}, a Julia package for polynomial optimization that supports non-commutative optimization, tracial polynomial optimization, trace polynomial optimization and state polynomial optimization.
\texttt{PCPOP} fully supports exact arithmetic computations and incorporates convenient functionalities such as algebraic reductions based on Gr\"obner basis methods, automatized symmetrization via Wedderburn decompositions, and Jordan algebra reductions. 
As a distinguished feature, \texttt{PCPOP} implements a specialized framework for polynomial computations in partially commutative variables that provides significant computational advantages for problems appearing in quantum information.

\end{abstract}

\tableofcontents*

\chapter{Introduction}
\label{ch:introduction} 

A polynomial optimization problem aims to find the maximal value of a polynomial over a region specified by polynomial constraints, and provides an extremely flexible framework to model a wide variety of problems that range from scientific research to industrial and technological applications. In his seminal work~\cite{lasserre2001global}, Lasserre proposed a hierarchy of semidefinite programs relaxing a polynomial optimization problem, whose optimal values converge (under mild assumptions) to the solution of the polynomial optimization problem. 
The framework of polynomial optimization has been further extended to the non-commutative setting in different forms \cite{pironio2010convergent, klep2022optimization, klep2024state}, becoming a central tool for researchers in quantum information, quantum foundations and quantum technologies.

The main limitation in practical applications is the lack of heuristic methods to choose informative relaxations for a target polynomial optimization problem. These relaxations are often obtained by a generic degree bound on the monomials, which makes their size grow exponentially and soon become intractable. In order to overcome these limitations, several techniques have been developed in the literature exploiting internal symmetries~\cite{gatermann2004symmetry, ioannou2021noncommutative}, Jordan algebra structure~\cite{permenter2020dimension, brosch2022jordan}, different forms of sparsity~\cite{klep2022sparse} and additional algebraic structure~\cite{pcpop} to reduce the computational cost of solving these semidefinite relaxations.
In particular, in \cite{pcpop} we propose a specialized framework for partially commutative polynomial optimization that exploits the algebraic structure induced by commutation relations (and other constraints) to boost polynomial computations and reduce the size of the semidefinite relaxations for polynomial optimization problems. 

\texttt{PCPOP} is a Julia package that automates constructing semidefinite programming relaxations for polynomial optimization problems. As a distinguishing feature, \texttt{PCPOP} implements a specialized framework for partially commutative polynomial optimization~\cite{pcpop} that is particularly effective for problems appearing in quantum information.
The package covers commutative polynomial optimization~\cite{lasserre2001global}, non-commutative polynomial optimization~\cite{pironio2010convergent}, tracial polynomial optimization~\cite{burgdorf2013tracial, burgdorf2016optimization}, trace polynomial optimization~\cite{klep2022optimization} and state polynomial optimization \cite{klep2024state}, offering a wide range of applications.
On top of the algebraic reductions exploiting the partial commutations, the package also supports more general algebraic reductions through Gr\"obner bases methods, automated symmetrization via Wedderburn decompositions~\cite{gatermann2004symmetry, ioannou2021noncommutative}, dimension reductions with Jordan algebras \cite{permenter2020dimension, brosch2022jordan} and fully supports exact arithmetic computations.

The aim of this manual is to offer an user-friendly guide on the package and its functionalities for a general audience through a rich and varied collection of examples. The theoretical background contained in the manual is brief and only introduces the indispensable terminology to explain the relevant frameworks and functionalities. We provide several references to the literature for further details.

\section{Semidefinite programs}
\label{sec:sdp}

A \emph{semidefinite program} in (complex) dimension $n$ has the form
\begin{align}
\max_X \ \ & \langle C, X \rangle
\label{eq:sdp_primal} \\
\textup{s.t.} \ \
& X \in Y + S \, , \nn \\
& X \geq 0  \, , \nn
\end{align}
for some (complex) hermitian matrices $C$ and $Y$ of size $n$ and a linear subspace $S$, which may be presented through a list of linear constraints $\langle A_i, X \rangle = b_i$. Semidefinite programs admit a \emph{dual} formulation that minimizes upper bounds to the primal problem, which is again a semidefinite program:
\begin{align}
\min_Z \ \ & -\innerprod{Y}{Z}
\label{eq:sdp_dual} \\
\st \ \
& Z \in C + S^\perp \, , \nn \\
& Z \geq 0  \, , \nn
\end{align}
where $S^\perp$ is the orthogonal complement of $S$.
Therefore, a dual feasible solution gives an upper bound to the primal problem and a primal feasible solution gives a lower bound for the dual problem. This relation is called \emph{weak duality}. In some cases, the optimal solutions of the primal and the dual problem coincide, a property called \emph{strong duality}. A sufficient condition for strong duality is that there exists an interior point in the (primal or dual) feasible region. Semidefinite programs admit polynomial time algorithms~\cite{karmarkar1984new, nesterov1994interior} and are extensively used for a wide class of applications~\cite{vandenberghe1996semidefinite}.

\section{Polynomial optimization}
\label{sec:relaxations}
\label{sec:pop}

For an alphabet $\vec x$, we denote $\langle \vec x \rangle$ (respectively $[\vec x]$) the free (commutative) monoid over $\vec x$, whose elements are free (commutative) words on the letters in $\vec x$. Elements of $\vec x$ are also called \emph{variables}, and elements of $\langle \vec x \rangle$  (respectively $[\vec x]$) are also called (commutative) \emph{monomials}. \emph{Complex polynomials} are finite linear combinations of monomials with coefficients in $\mathds C$, whose collection we denote $\mathds C\langle \vec x \rangle$. An \emph{involution} on $\vec x$ is a map that satisfies $(x^*)^* = x$ for each letter $x \in \vec x$, which extends to monomials $w = x_1\ldots x_n$ via $w^* = x_n^* \ldots x_1^*$. An involution linearly extends to polynomials acting as complex conjugation on the coefficients.
A polynomial optimization problem over $\mathds C\langle \vec x \rangle$ is specified by a triple $(p, R, S)$, where $p \in \mathds C\langle \vec x \rangle$ is an hermitian objective polynomial, $R \subset \mathds C\langle \vec x \rangle$ is a collection of equality constraints and $S\subset \mathds C \langle \vec x \rangle$ a collection of hermitian inequality constraints. We display the problem as
\begin{align}
    p^* = \sup \quad & p(\vec x)
    \label{eq:ncpop} \\
    \operatorname{s.t.} \quad & r(\vec x) = 0 & r \in R \, , \nonumber \\
    & s(\vec x) \geq 0 & s \in S \, . \nonumber
\end{align}
A \emph{solution} is a linear functional $L : \mathds C\langle \vec x \rangle \to \mathds C$ that satisfies $L(\id) = 1$ and $L(q^*sq) \geq 0$ for each $s \in \{\id\} \cup S \cup \pm R$ and $q \in \mathds C\langle \vec x \rangle$, which has \emph{value} $L(p) \in \mathds R$. When $(p, R, S)$ only involve polynomials with real coefficients, it is enough to consider real solutions $L:\mathds R\langle \vec x \rangle \to \mathds R$.
The \emph{optimal value} $p^*$ is the supremum of the values over all possible solutions. Namely,
\begin{align}
    p^* = \sup \quad & L(p) 
    \label{eq:ncpop_moments}\\
    \operatorname{s.t.} \quad & L(\id) = 1 \nonumber \\
    & L(q^*s q) \geq 0 & s \in \{\id\} \cup S \cup \pm R \, . \nonumber
\end{align}
We call $L(w)$ the \emph{moment} of the monomial $w \in \langle \vec x \rangle$, since in the commutative setting $L(w)$ is precisely the moment of $w$ with respect to some probability measure. 
Under a boundedness assumption, solutions of Problem~\eqref{eq:ncpop_moments} coincide exactly with the expectation values under a state of bounded operators satisfying the constraints in $R$ and $S$ \cite{pironio2010convergent}. Namely, there are operators $X_1, \ldots, X_n \in \mathcal B(H)$ and a state $\rho \in \mathcal S(H)$ over some Hilbert space $H$ such that $r(\vec X) = 0$ for each $r \in R$, $s(\vec X) \geq 0$ for each $s \in S$ and $L(q) = \rho(q(\vec X))$ for each $q \in \mathds C \langle \vec x \rangle$.

The restriction of Problem~\eqref{eq:ncpop_moments} to a finite subspace is a semidefinite program that provides a relaxation of the original problem, often called the \emph{moment relaxation}. Indeed, let $B_d$ be the collection of monomials with degree no greater than $d$ and consider the subspace $V_d = \operatorname{span} \{u^* v : u, v \in B_d\}$. The \emph{moment matrix} localized at $s \in \{\id\}\cup S \cup \pm R$ has entries $M_s(u,v) = L(u^*sv)$ for $u, v$ in a suitable subset of $B_d$ such that $u^*sv \in V_d$ (alternatively, extend $V_d$ with those new monomials $u^*sv$). For a polynomial $p= \sum_w p_w w$ we have
\begin{align}
    p^*_d = \sup \quad & \sum_w p_w L(w) \label{eq:sdp_relax} \\
    \operatorname{s.t.} \quad & L(\id) = 1 \nonumber \\
    & M_s \geq 0 & s \in \{\id\} \cup S \cup \pm R \, . \nonumber
\end{align}
It is immediate that $p^* \leq p^*_d$ for each $d$, since a solution of Problem~\eqref{eq:ncpop_moments} provides by restriction a solution of Problem~\eqref{eq:sdp_relax}. Moreover, (under a boundedness assumption) the sequence of upper bounds $p^*_d$ converges to the optimal value $p^*$ \cite{pironio2010convergent}. 

Equivalently, via the Positivstellensatz for non-commutative polynomials \cite{helton2004positivstellensatz}, the dual of Problem~\eqref{eq:ncpop_moments} can be understood as optimizing over \emph{sum of squares} decompositions:
\begin{align}
    p^* = \inf \quad & t 
    \label{eq:ncpop_sos} \\
    \operatorname{s.t.} \quad & t - p \in \operatorname{SOS}(S, R)\, . \nonumber
\end{align}
Here, $\operatorname{SOS}(S, R)$ is the collection of polynomials $\sum_i q_i^* s_i q_i$ where $s_i \in \{\id\}\cup S \cup \pm R$ and $q_i \in \mathds C\langle \vec x \rangle$. The restriction of the sum of squares decomposition to polynomials $q_i$ in a finite subspace $V \subset \mathds C\langle \vec x \rangle$ gives a semidefinite program. The corresponding program for the subspace $V_d$ of polynomials with degree no greater than $d$ is called the \emph{sum of squares relaxation} of level $d$, which is the dual of the \emph{moment relaxation} in Equation~\eqref{eq:sdp_relax}. Namely,
\begin{align}
    p^*_d = \inf \quad & t 
    \label{eq:sdp_sos_relax} \\
    \operatorname{s.t.} \quad  & N_s \geq 0 & s \in \{\id\} \cup S \cup \pm R \, , \nonumber \\
    & (t - p)_m = \textstyle \sum_{a^*s_i b = m} c_i N_{s}(a, b)  & m \in \langle \vec x \rangle \, . \nonumber
\end{align}
In the last condition, $(t- p)_m$ denotes the coefficient of the monomial $m$ in the polynomial $t - p$, and the sum runs over all $s = \sum_i c_i s_i \in \{\id\} \cup S \cup \pm R$ with $a^*s_ib = m$. 

\section{Tracial polynomial optimization}
\label{sec:pop_tracial}

In a \emph{tracial} polynomial optimization problem \cite{burgdorf2013tracial}, one optimizes over tracial linear functionals, that is, $L(uv) = L(vu)$ for each pair of monomials $u, v \in \langle \vec x \rangle$. The moment relaxations for tracial optimization have the form
\begin{align}
    p^*_d = \sup \quad & \sum_w p_w L(w) \label{eq:tr_sdp_relax} \\
    \operatorname{s.t.} \quad & L(\id) = 1 \nonumber \\
    & M_s \geq 0 & s \in \{\id\} \cup S \cup \pm R \, , \nonumber \\
    & L(uv) = L(vu) \, . \nonumber
\end{align}
Under boundedness assumptions, feasible solutions can be represented as expectation values under a tracial state of bounded operators satisfying the constraints $S$ and $R$ \cite[\S 3.2.2]{burgdorf2013tracial}.
The dual of Problem~\ref{eq:tr_sdp_relax} corresponds with a sum of squares decomposition up to cyclic equivalence.
Although not every polynomial with non-negative trace is cyclically equivalent to a sum of squares \cite[\S 2.3]{burgdorf2013tracial}, under certain flatness conditions these relaxations are exact for trace polynomial optimization \cite[Theorem 3.12]{burgdorf2013tracial}. Namely, there exist operators $X_1, \ldots, X_n \in \mathcal B(H)$ in some finite dimensional Hilbert space $H$ satisfying $r(\vec X) = 0$ and $s(\vec X)\geq 0$ for each $r \in R$ and $s \in S$ such that $ L(q) = \tr(q(\vec X))$ for each $q \in \mathds C \langle \vec x \rangle$.

\section{State polynomial optimization}
\label{sec:spop}

The frameworks of trace and state polynomials~\cite{klep2022optimization, klep2024state} extend the scope of the polynomial optimization problems discussed in the previous sections, by accommodating non-linear expressions on the moments. These cover several additional interesting problems in quantum information.
For this, consider the alphabet $\rho \langle \vec x \rangle = \{\rho(w) : w \in \langle \vec x \rangle\}$ of free words over $\vec x$ under a new \emph{state symbol} $\rho$. \emph{State polynomials} are elements of $\mathds C\langle \vec x \rangle \times [\rho \langle \vec x \rangle]$, which have the form 
\begin{equation}
p = w_0 \rho(w_1) \ldots \rho(w_n)
\end{equation}
for certain $w_0, \ldots, w_n \in \langle \vec x \rangle$. We denote $\rho : \mathds C\langle \vec x \rangle \times [\rho \langle \vec x \rangle] \to \mathds C [\rho \langle \vec x \rangle]$ the \emph{state projection} induced via $\rho(w_0 \rho(w_1) \ldots \rho(w_n)) = \rho(w_0) \rho(w_1) \ldots \rho(w_n)$. Feasible solutions of state polynomial optimization problems are unital positive linear functionals $L : \mathds C \langle \vec x \rangle \times [\rho \langle \vec x \rangle] \to k$ that satisfy the equality and inequality constraints and additionally interpret the state symbol. Namely, $L(q^* s q) \geq 0$ and $L(p) = L(\rho(p))$ for each $s \in \{\id\}\cup S \cup \pm R$ and $p, q \in \mathds C\langle \vec x \rangle$.
State polynomial optimization problems admit moment relaxations as Equation~\eqref{eq:sdp_relax}, with additional linear constraints to interpret the state symbol \cite[Equation 6.6]{klep2024state}:
\begin{align}
    p^*_d = \sup \quad & \sum_w p_w L(w) \label{eq:state_sdp_relax} \\
    \operatorname{s.t.} \quad & L(\id) = 1 \nonumber \\
    & M_s \geq 0 & s \in \{\id\} \cup S \cup \pm R \, , \nonumber \\
        & L(a) = L(\rho(a)) \, . \nonumber
\end{align}
Therefore, the values $p_d^*$ give a convergent hierarchy of upper bounds to the optimal value $p^*$ of the state polynomial optimization problem. Under a boundedness condition that is satisfied in the applications we consider, feasible solutions can be represented with bounded operators and a state that interprets the state symbol \cite[Theorem 5.5]{klep2024state}. Namely, there are bounded operators $X_1, \ldots, X_n \in \mathcal B(H)$ and a state $\lambda \in \mathcal S(H)$ over some Hilbert space $H$ such that $r(\vec X) = 0$ for each $r \in R$, $s(\vec X) \geq 0$ for each $s \in S$ and $L(q) = \lambda(q(\vec X))$ for each $q \in \mathds C \langle \vec x \rangle$, where we evaluate each state monomial $w_0 \rho(w_1)\ldots \rho(\omega_n)$ as $w_0(\vec X) \lambda(w_1(\vec X)) \ldots \lambda (w_n (\vec X))$.

\section{Trace polynomial optimization}
\label{sec:tracepop}

For \emph{trace polynomials} we denote the state symbol by $\tau$ and we identify $\tau (uv) = \tau(vu)$ for each pair of free words $u, v \in \langle \vec x \rangle$. Feasible solutions to trace polynomial optimization problems are linear functionals $L$ that satisfy the equality and inequality constraints, interpret the tracial state symbol. Namely, $L(a^*sa)\geq 0$ and $L(a) = L(\tau(a))$ for every $s \in \{\id\}\cup S \cup \pm R$ and trace monomial $a$. This automatically forces $L$ to be a tracial state, i.e. $L(ab) = L(ba)$ for each trace monomials $a$ and $b$. The moment relaxations have the form  \cite[Equation 5.17]{klep2022optimization}
\begin{align}
    p^*_d = \sup \quad & \sum_w p_w L(w) \label{eq:trace_poly_sdp_relax} \\
    \operatorname{s.t.} \quad & L(\id) = 1 \nonumber \\
    & M_s \geq 0 & s \in \{\id\} \cup S \cup \pm R \, , \nonumber \\
        & L(a) = L(\tau(a)) \, . \nonumber
\end{align}
Under mild assumptions, feasible solutions to trace polynomial optimization problems can be represented with operators in a von Neumann algebra with a tracial state that interprets the tracial state symbol \cite[Theorem 4.4]{klep2022optimization}. Namely, there are bounded operators $X_1, \ldots, X_n \in \mathcal A$  over some von Neumann algebra $\mathcal A$ with tracial state $\lambda \in \mathcal S(\mathcal A)$ such that $r(\vec X) = 0$ for each $r \in R$, $s(\vec X) \geq 0$ for each $s \in S$ and $L(q) = \lambda(q(\vec X))$ for each $q \in \mathds C \langle \vec x \rangle$, where we evaluate each trace monomial $w_0 \tau(w_1)\ldots \tau(\omega_n)$ as $w_0(\vec X) \lambda(w_1(\vec X)) \ldots \lambda (w_n (\vec X))$.

\section{Implementation of the semidefinite relaxations}

There are different approaches to encode the constraints in a semidefinite program. Although theoretically equivalent, the choice for the implementation can have drastic consequences in the performance of the solver and the numerical stability of the algorithms. The best implementation generically depends both on the problem and the technical details of the solver, and most solvers preprocess the problem to optimize its efficacy.

\texttt{PCPOP} offers three different implementations for the semidefinite programs:
\begin{enumerate}[nosep]
    \item \textbf{Scalar variables}. We create a scalar variable $x_w$ for the moment $L(w)$ of each monomial $w$ in the relaxation. Then, we build matrices $X_s(x)(u, v) = x_{u^*v}$ using these scalar variables, that correspond with the moment matrices $M_s$, and impose positive semidefinite constraints. Namely,
    \begin{align}
    \sup \quad & \langle c, x \rangle \label{eq:sdp_scalar} \\
    \st \quad & X_s(x) \geq 0 \, , \nonumber \\
    & A(x) \geq a \, , \nonumber \\
    & B(x) = b \, . \nonumber
    \end{align}
    This is implemented in \texttt{pcpop} when \texttt{primal=true} and \texttt{canonical=false}.
    \item \textbf{Matrix variables}. We create positive semidefinite matrix variables $X_s$ for the moment matrices $M_s$. Then, we impose linear constraints $E_i(X) = 0$ on their entries to capture the linear dependencies between the monomials in the moment matrices. Despite involving seemingly more variables and constraints, this implementation behaves better in most of the applications we have considered. Therefore, this is used by default both for the moment and sum of squares relaxations.
    \begin{align}
    \textstyle \sup \quad & \sum_s \langle C_s, X_s \rangle \label{eq:sdp_matrix} \\
    \st \quad & X_s \geq 0 \, , \nonumber \\
    & A(X) \geq a \, , \nonumber \\
    & B(X) = b \, , \nonumber \\
    & E_i(X) = 0 \, . \nonumber
\end{align}
This is the default implementation in \texttt{pcpop} when \texttt{primal=true} or \texttt{primal=false}.
    \item \textbf{Canonical form}. We introduce positive semidefinite slack variables $Z$ to convert all inequality constraints $A(X) \geq a$ into equalities $A(X) - Z = a$. Then, we combine all positive semidefinite matrix variables into one positive semidefinite matrix variable $X$ with zeros outside the diagonal blocks. Therefore, the semidefinite program only involves one positive semidefinite matrix variable and equality constraints,
    \begin{align}
    \sup \quad & \langle C, X \rangle \label{eq:sdp_vector} \\ 
    \st \quad & X \geq 0 \, , \nonumber \\
    & A(X) = a \, , \nonumber \\
    & B(X) = b \, . \nonumber
    \end{align}
    This is employed (in vectorized form) during the Jordan algebra reductions to find invariant subspace, implemented in \texttt{pcpop} when \texttt{reduce=true}.
\end{enumerate}
Notice that changing the objective function $C$ and the linear constraints $A$ and $B$ from one implementation to another only requires elementary computations. We remark that the implementation with scalar variables already provides an invariant subspace which has a similar size to the one obtained with Jordan reduction for most examples we considered.
\chapter{Features of \texttt{PCPOP}}
\label{sec:features}

\texttt{PCPOP} offers a framework for polynomial computations with different functionalities. The main functionality is building and solving semidefinite programming relaxations for polynomial optimization problems, and the major novel contribution is the implementation of the recently developed framework of partially commutative polynomial optimization \cite{pcpop}. \texttt{PCPOP} supports non-commutative polynomial optimization \cite{lasserre2001global, pironio2010convergent} [Section~\ref{sec:pop}], tracial polynomial optimization \cite{burgdorf2013tracial, burgdorf2016optimization} [Section~\ref{sec:pop_tracial}], state polynomial optimization \cite{klep2024state} [Section~\ref{sec:spop}] and trace polynomial optimization \cite{klep2022optimization} [Section~\ref{sec:tracepop}].
As additional functionalities, \texttt{PCPOP} implements different algebraic reductions based on Gr\"obner basis techniques and specialized representations exploiting the partial commutations~\cite{pcpop}, symmetry reductions and Jordan algebra reductions~\cite{gatermann2004symmetry, permenter2020dimension}, which we briefly explain in this chapter.

\section{Partially commutative polynomial computations}
\label{sec:pcwords}

The most distinguishing feature of the framework for partially commutative polynomial optimization~\cite{pcpop} is an effective representation of polynomials in partially commutative letters that automatically provides canonical forms for special classes of constraints ubiquitous in quantum information (projections, unitaries, unipotents and orthogonality). Instead of representing words as one-dimensional sequences of letters and treating strings as equivalent when two commuting letters appear in different order, \texttt{PCPOP} implements canonical forms for these equivalence classes that allow to perform algebraic computations. In the extreme case where all variables commute, these normal forms simply count the number of occurrences of each letter in a word (often called \emph{exponents}), recovering the more effective exponent representation of commutative polynomials.

Let $\mathds C \langle \vec x \rangle$ be a polynomial ring on the partially commutative variables $\vec x$. The dependencies among the variables are described with the \emph{dependence relation} $D \subset \langle \vec x \rangle \times \langle \vec x \rangle$, which we assume to be reflexive for convenience. That is, $(x, y) \in D$ when $x$ and $y$ are the same or do not commute. The pair $G = (\vec x, D)$ is called the \emph{dependence graph} of the partially commutative variables $\vec x$. A \emph{clique} in $G$ is a set $C \subset \vec x$ of pairwise non-commutative variables, that is, $C \times C \subset D$. Let $\mathcal C = (C_1, \ldots, C_n)$ denote the collection of all maximal cliques in $G$. The homomorphism $\pi_{i} : \langle \vec x \rangle \to \langle C_i \rangle$ induced via
\begin{equation}
    \pi_i(x) = \left\{
    \begin{array}{ll}
    x & x \in C_i \\
    1 & x \not \in C_i
    \end{array}\right.
\end{equation}
projects a word $w \in \langle \vec x \rangle$ in partially commutative letters to the (fully non-commutative) subword $w_i = \pi_i(w)$ containing only the letters in $w$ that belong to the clique $C_i$, in their order of appearance. The original word $w$ can be recovered from the collection $(w_1, \ldots, w_n)$ of its clique projections, and two words equivalent up to partial commutations produce the same clique projections~\cite{duboc1986some}. Therefore, the collection of clique projections provides a canonical form of partially commutative words, called the \emph{clique representation}. Other canonical forms are discussed in~\cite{pcpop} based on the seminal works~\cite{cartier1969applications, mazurkiewicz1977concurrent, diekert1997handbook}. The clique representation is specially suitable for our computational implementation, since it easily allows to perform all essential algebraic computations required for polynomial optimization.

Fix an enumeration $\mathcal C=(C_1, \ldots, C_n)$ of the maximal cliques in the dependence graph $G=(\vec x, D)$ of a partially commutative alphabet. Let $u$ and $v$ be partially commutative words with clique representations $(u_1, \ldots, u_n)$ and $(v_1, \ldots, v_n)$.
\begin{enumerate}
\item \textbf{Multiplication.}
 The clique representation of the product $uv$ is $(u_1 v_1, \ldots, u_n v_n)$.
\item \textbf{Involution}. Notice that under the assumption that the involution induces an automorphism of the dependence graph (i.e. $(x^*, y^*)\in D$ if and only if $(x, y) \in D$), the involution transforms maximal cliques into maximal cliques. That is,  $C_i^* = C_{i^*}$ for certain index $i^*$.
Then, the clique representation of $u^*$ is $(u_{1^*}^*, \ldots, u_{n^*}^*)$. 

\item \textbf{Equality}. The words $u$ and $v$ are equivalent up to partial commutations if and only if their clique representations coincide, $(u_1, \ldots, u_n) = (v_1, \ldots, v_n)$.

\item \textbf{Division.} 
The word $u$ divides $v$ if $l u r = v$ for some partially commutative words $l$ and $r$. Divisibility can be decided with the clique representations. First, we find all pairs of tuples $(l_1, \ldots, l_m)$ and $(r_1, \ldots, r_m)$ such that $l_k u_k r_k = v_k$ as non-commutative words, which is an instance of substring matching problem. Then, we check if any such pair $(l_1, \ldots, l_m)$ and  $(r_1, \ldots, r_m)$ actually corresponds with a partially commutative words. This can be done constructing a \emph{graph occurrence representation}~\cite{duboc1986some} and checking that there are no cycles.

\item \textbf{Tracial equivalence}.
In the frameworks of tracial and trace polynomials, one essential ingredient is to identify words that differ in a cyclic permutation of its letters. 
The words $u$ and $v$ are equivalent up to cyclic permutations if there exists a sequence of partially commutative words $u_0,u_1 \ldots u_n$, such that $u_0=u$, $u_n=v$ and for consecutive words $u_i=t_is_i$ and $u_{i+1}=s_it_i$ for some partially commutative words $s_i$ and $t_i$. 
Notice that this condition is not a polynomial constraint, therefore Gr\"obner bases methods cannot be used to obtain canonical forms for these equivalence classes. 
\texttt{PCPOP} implements an algorithm introduced in \cite{liu1990efficient} that decides in linear time when two partially commutative words are equivalent up to cyclic permutations.
Namely, $u$ is equivalent to $v$ up to cyclic permutations if and only if both $u$ and $v$ have the same exponents and $u$ divides $v^n$ for some $n \leq |\vec x|$.
\end{enumerate}

For the purpose of illustrating these computations with a concrete example, consider the 
partially commutative alphabet over the hermitian letters $\vec x = \{a, b, c, d, e\}$ with maximal cliques $\mathcal C = (ab, bc, cd, de, ea)$. The clique representation of the partially commutative words $u=bac$, $v = ab$ and $w = ecadeadbc$ are $(ba, bc, c, \one, a)$, $(ab,b, \one, \one, a)$ and $(aab, cbc, cddc, eded, eaea)$. The clique representation of $uv = bacab$ is $(baab, bcb, c, \one, aa)$. Since the alphabet is hermitian, the maximal cliques are fixed by under the involution and the clique representation of $u^*$ is simply $(ab, cb, c, \one, a)$. The word $v$ does not divide $u$ but it divides $w$, as witnessed by $(a, c, cdd, eded, eae)(ab, b, \one, \one, a)(\id, c, c, \id, \id) = (aab, cbc, cddc, eded, eaea)$ and the fact that $(a, c, cdd, eded, eae)$ and $(\id, c, c, \id, \id)$ are clique representations for the words $l = eacded$ and $r = c^2$, which can be obtained constructing the graph occurrences. Last, $u$ is equivalent to $acb$ and $cba$ up to cyclic permutations, but not to the words $abc$, $bca$ or $cab$ with the same exponents since $u$ does not divide any power of these words (for non-commutative words second power is enough).


\section{Graph products}
\label{sec:graph_product}

We can extend the notion of dependence among letters to dependence among algebras. Let $\vec A = (A_1, \ldots, A_n)$ be a tuple of $n$ (partially commutative) polynomial algebras, and a \emph{dependence relation} $D\subset \vec A \times \vec A$, which is assumed to be reflexive. The \emph{graph product} of the algebras $\vec A$ with respect to the dependence graph $G=(\vec A, D)$ is the partially commutative algebra generated by $A_1, \ldots A_n$ with the internal relations inside each local algebra $A_i$ plus the additional commutation relations $a_i a_j = a_j a_i$ between elements $a_i \in A_i$ and $a_j \in A_j$ of \emph{independent} local algebras, $(A_i, A_j) \not\in D$. We can understand this graph product construction of polynomial algebras as a short-cut to describe partially commutative polynomial algebras for which collections of letters share the same commutation relations. In particular, every partially commutative polynomial algebra can be obtained with the graph product construction of polynomial algebras with one variable. Moreover, every partially commutative polynomial has a universal graph product constructions \cite{pcpop}, which has a special physical significance: it provides an algebraic notion of \emph{subsystems} based on commutation classes.
Several properties of graph product constructions for monoids and polynomial algebras have been considered in the literature \cite{da2001graph, atecs2011grobner, dandan2023graph}.
For instance, the graph product of monoids with decidable word problem, also has decidable word problem \cite[Theorem 6.5]{da2001graph}.
This means that a collection of Gr\"obner bases $(G_1, \ldots, G_n)$ for local binomial constraints $(P_1, \ldots, P_n)$, i.e. each $p_i \in P_i \in A_i$ has the form $p_i = m_i - n_i$ for some monomials $m_i$ and $n_i$, can be raised to a global Gr\"obner basis in the graph product. 

\section{Algebraic reductions}

In standard implementations of polynomial optimization problems, equality constraints are imposed over the moments in the semidefinite programming relaxations. From a practical point of view, it is more efficient to start from the beginning with canonical forms for the equivalence classes induced by the equality constraints. This reduces both the number of variables and constraints in the semidefinite programming relaxations. 

More precisely, a collection $R$ of equality constraints induces an equivalent relation between polynomials, with equivalence classes $[p] = \{ q : p - q \in \langle R \rangle \}$. Therefore, instead of considering the polynomial optimization problem $(p, R, S)$ [Eq.~\eqref{eq:ncpop}] and imposing the constraints in $R$, we can directly consider the reduced polynomial optimization problem $([p], \emptyset, [S])$ over the equivalence classes induced by $R$:
    \begin{align}
    p^* = \sup \quad & [p](\vec x) \label{eq:pop_algebraic} \\
    \operatorname{s.t.} \quad & [s](\vec x) \geq 0 & s \in S \, . \nonumber
    \end{align}
    That is, we consider linear functionals $L$ over equivalence classes, which reduces both the number of variables and constraints in the semidefinite programming relaxations [Eq.~\eqref{eq:sdp_relax}].
    Indeed, the moment matrices decompose as $ M_s = \sum_{[w]} L([w]) M_{s,[w]}$, where $L([w])$ is the scalar variable corresponding to the monomials with canonical form $[w]$ and $M_{s, [w]}$ is the matrix of occurrences of monomials with canonical form $[w]$ in $M_s$. The semidefinite relaxation can be written as
    \begin{align}
    p^*_d = \sup \quad & \sum_{[w]} p_{[w]} L([w]) \label{eq:sdp_alg_reduction} \\
    \operatorname{s.t.} \quad & L([\id]) = 1 \, , \nonumber \\
    & M_s = \sum_{[w]} L([w]) M_{s,[w]} \geq 0 & s \in \{\id\} \cup S \, . \nonumber
    \end{align}   
    Canonical representations for these equivalence classes can be obtained with Gr\"obner bases methods~\cite{mora1994introduction, scala2009letterplace, xiu2012non}. \texttt{PCPOP} provides Gr\"obner bases computations to obtain canonical representations using \texttt{AbstractAlgebra}, which relies on the non-commutative version of Buchberger algorithm proposed in \cite{xiu2012non}.
    
    Despite theoretically providing canonical representations for arbitrary equality constraints, Gr\"obner basis computations may not terminate in the non-commutative setting \cite{mora1994introduction}, and even finite truncations can become expensive to compute. As an alternative,
    \texttt{PCPOP} implements specialized canonical forms for polynomials in partially commutative letters \cite{pcpop} that support additional constraints such as projections, unitaries, unipotents and orthogonality.
    Further constraints can be imposed over the moments on the semidefinite programming relaxations.
    This approach proves to be specially effective in problems arising in quantum information. In Chapter~\ref{ch:benchmarking}, we show problems for which \texttt{PCPOP} offers an advantage against other state-of-the-art packages for (non-commutative) polynomial optimization relying on Gr\"obner-like reductions.

\section{Symmetry reductions}
\label{sec:symmetries}

Another general method to reduce the size of a semidefinite program and speed up the computation is to exploit internal symmetries \cite{gatermann2004symmetry}. We say that a unitary matrix $U$ is a \emph{symmetry} for a semidefinite program [Eq.~\eqref{eq:sdp_primal}] if it leaves the feasible region and objective function invariant. That is, $UCU^* = C$ and $U(Y+S)U^* = Y + S$. Symmetries form a group under composition. 
    Notice that if $X$ is an optimal solution and $U$ a symmetry, then $UXU^*$ is also an optimal solution. Therefore, we can restrict the optimization to the symmetric subspace $X = UXU^*$, which admits a block-diagonal decomposition $X = \bigoplus_i X_i$ by  Wedderburn classification theorem for matrix algebras. This reduces the positivity constraint to smaller blocks
    \begin{align}
    \max_{X} \ \ & \langle \bigoplus C_{i}, \bigoplus X_i \rangle \\
    \textup{s.t.} \ \
    & X_i \in Y_i + S_i \, , \nn \\
    & X_i \geq 0  \, . \nn 
    \end{align}  
    
    When a polynomial optimization problem $(p, R, S)$ is invariant under some transformation, the semidefinite relaxations [Eq.~\eqref{eq:sdp_relax}] will manifest symmetries that can be exploited to reduce their computational cost~\cite{ioannou2021noncommutative}. Given a collection of automorphisms on the algebra of polynomials that leave a polynomial optimization problem invariant, \texttt{PCPOP} supports automated symmetry reductions of the corresponding semidefinite programming relaxations. The symmetrization is based on Wedderburn decompositions using \texttt{SymbolicWedderburn} \cite{kaluba2019aut}.

\section{Jordan algebra reductions}

An alternative axiomatic approach to reduce semidefinite programs was introduced in \cite{permenter2020dimension}, characterizing those feasible subspaces that preserve optimal solutions. A linear space $V$ of positive semidefinite matrices is \emph{invariant} for the primal-dual pair in Eqs.~\eqref{eq:sdp_primal} and~\eqref{eq:sdp_dual} if it preserves positivity, primal feasibility and dual feasibility. That is, 
    $(i)$ $P_V(A) \geq 0$ for all $A\geq 0$, $(ii)$ $P_V(Y + S) \subset Y + S$, and $(iii)$ $P_V(C + S^\perp) \subset C + S^\perp$; where $P_V$ denotes the orthogonal projection to $V$.
   The optimal value of the semidefinite program is preserved in the invariant subspace:
\begin{align}
\max \ \ & \langle P_V(C), X \rangle
\label{eq:primal-reduced} \\
\textup{s.t.} \ \
& X \in P_V(Y) + S \cap V \nn \\
& X \geq 0 \, , \nn \\
\min \ \ & - \langle P_V(Y), Z \rangle
\label{eq:dual-reduced} \\
\textup{s.t.} \ \
& Z \in P_V(C) + S^\perp\cap V \nn \\
& Z \geq 0 \, . \nn
\end{align}
The conditions above are equivalent to $P_S(C)\in V$, $P_{S^\perp}(Y) \in V$, $P_S(V) \subset V$ and $\{A^2 : A \in V\} \subset V$, therefore a \emph{minimal} invariant subspace can be obtained algorithmically \cite[Theorem 3.2]{permenter2020dimension}. Moreover, there exist efficient combinatorial relaxations that are at least as good as any symmetry reduction \cite[\S 5]{permenter2020dimension}, with the advantage that the symmetries do not need to be explicitly provided. 

\texttt{PCPOP} offers a functionality to obtain invariant subspaces in the semidefinite relaxations of polynomial optimization problems using \texttt{SDPSymmetryReduction} \cite{brosch2022jordan}, which relies on a randomized implementation of the combinatorial relaxation proposed in \cite[\S 5]{permenter2020dimension}. Moreover, the corresponding invariant subspace can be numerically block-diagonalized using a randomized algorithm \cite[Algorithm 4.1]{murota2007numerical}.
Notice that the algebraic reductions in Equation~\eqref{eq:pop_algebraic} using canonical forms automatically provide invariant subspaces for the semidefinite programming relaxations of polynomial optimization problems. In most of the examples from quantum information that we consider, there is no significant gain through Jordan algebra reductions.

\section{Exact arithmetic} 

Algebraic computations in \texttt{PCPOP} support exact arithmetic. Additionally, Gr\"obner bases computations with \texttt{AbstractAlgebra} and the Wedderburn decomposition used for the symmetry reduction of semidefinite programs implemented through \texttt{SymbolicWedderburn} support exact arithmetic. Solutions of semidefinite programs can be rounded to exact arithmetic using methods implemented in \texttt{ClusteredLowRankSolver} \cite{leijenhorst2024solving}.

\section{Comparison with other packages}
\label{sec:comparison}

We compare the features of \texttt{PCPOP} with other packages supporting non-commutative polynomial optimization. 

\begin{enumerate}[]
    \item \texttt{Ncpol2sdpa} Python package \cite{ncpol2sdpa, wittek2015algorithm}: one of the first packages supporting non-commutative polynomial optimization. It allows to relax equality constraints with substitution rules that reduce the number of monomials and hence the size of the relaxations. These substitutions rules, however, may fail to identify polynomials in the same equivalence class even for constraints only involving partial commutations.
    \texttt{Ncpol2sdpa} does not support tracial, trace or state polynomials.
    \item \texttt{QuantumNPA.jl} Julia package \cite{quantumnpa}: implementation of the semidefinite programming hierarchies for non-commutative polynomial optimization with some functionalities tailored to quantum information problems. It provides effective representations for the commutation relations in Bell scenarios. Some special constraints in quantum information problems are imposed during the polynomial computations. Further equality constraints are imposed with conditions on the moments. It does not support tracial, trace or state polynomials.
    \item \texttt{NCTSSOS} Julia package \cite{NCTSSOS, magron2022sparse}: a modern package that incorporates several sparsity reduction tools for non-commutative polynomial optimization. These sparsity reductions may provide informative approximations for big problems that can not be solved otherwise. Although these sparsity techniques can in principle be specialized to the partially commutative setting, \texttt{PCPOP} does not currently implement sparsity reductions and \texttt{NCTSSOS} does not currently implement partially commutative computations. \texttt{NCTSSOS} supports tracial, trace and state polynomial optimization. 
    \item \texttt{SumOfSquares.jl} Julia package \cite{weisser2019polynomial, legat2017sos}: transforms polynomial optimization problems into sum of squares decomposition problems. It supports Gr\"obner bases computations and general coefficient rings, and it incorporates several symmetry and sparsity reductions. It does not directly support tracial, trace or state polynomials.
    \item \texttt{Inflation} Python package \cite{inflation}: implements semidefinite programming relaxations for non-commutative polynomial optimization problems. \texttt{Inflation} automates the semidefinite programming relaxations obtained with inflation techniques \cite{wolfe2019inflation, wolfe2021quantum} for polynomial optimization problems over causal networks, which are not currently implemented in \texttt{PCPOP}. \texttt{Inflation} does not support tracial, trace or state polynomials.
    \item \texttt{Moments} C$++$ implementation with Matlab interface \cite{garner2024introducing}: implements semidefinite programming relaxations for non-commutative polynomial optimization problems. \texttt{Moment} supports algebraic reductions exploiting the equality constraints and symmetry reductions. It provides additional functionalities for certain problems in quantum information including Bell scenarios, cryptographic protocols and inflation hierarchies. \texttt{Moment} does not support tracial, trace or state polynomials.
\end{enumerate}

The performance of \texttt{PCPOP} is benchmarked against \texttt{Ncpol2sdpa}, \texttt{QuantumNPA} and \texttt{Moment} in different scenarios. The details can be found in Chapter~\ref{ch:benchmarking}. 


\chapter{Implementation}
\label{ch:implementation}

In this chapter we briefly discuss some technical details about the actual implementation, which are not necessary to follow the tutorial and applications in Chapters~\ref{ch:tutorial} and \ref{ch:applications}.

\section{Non commutative monoid}

Non-commutative monoids and non-commutative monomials are implemented using the data structures $\texttt{NCMonoid}$ and $\texttt{NCWord}$. These are wrappers around the data structures in the non-commutative algebra package \texttt{AbstractAlgebra}, which handles the arithmetic of non-commutative monomials and reduction algorithms based on Gr\"obner basis computations implemented in \texttt{PCPOP}.

\section{Partially commutative monoid}

Partially commutative monoids and partially commutative monomials are implemented using \texttt{GraphProductMonoid} and \texttt{GraphProductWord} data structures. Arithmetical computations with partially commutative polynomials are handled by default with clique representations.

     \textbf{Clique representation of partially commutative words}. These are stored as an array of arrays in the field \texttt{clique\_words} of the  data structure \texttt{PCMonomial}. The $i$-th array in \texttt{clique\_words} stores the projection of the word to the $i$-th clique. Two additional fields \texttt{edge\_l} and \texttt{edge\_r} store the \emph{initial} and \emph{final} letters in the word (i.e. that can be moved to first and last place with partial commutations). Storing these fields makes monomial multiplication more efficient, as stressed in \cite{pcpop}.

\textbf{Multiplication of partially commutative words.} Given two partially commutative monomials $u$ and $v$, the product $uv$ is obtained by concatenating element-wise the clique words of $u$ and $v$. Then, we evaluate \texttt{edge\_l} and \texttt{edge\_r} of the obtained \texttt{clique\_words} for $uv$ to create the product \texttt{PCMonomial}. We take care of additional constraints such as \emph{projectors, unitary, unipotent} implicitly in the multiplication process. For this, we sequentially go through pairs $(r, l)$ of letters $r$ in \texttt{edge\_r} of $u$ and $l$ in \texttt{edge\_l} of $v$ such that $rl$ appears in one of the internal constraints. Notice that each letter appears in at most one of the internal constraints, so there is no choice involved in the procedure. We reduce each $rl$ and store the product of the reductions for all such pairs $(l,r)$ into a new monomial $m$. Then each $r$ is deleted from the \texttt{edge\_r} and the last occurrence of $r$ is deleted from the \texttt{clique\_words} of $u$, obtaining a new $u'$. Similarly, each $l$ is deleted from \texttt{edge\_l} and the first occurrence of $l$ is deleted from \texttt{clique\_words} of $v$, obtaining a new $v'$. Then, multiply $u' m v'$. Therefore, multiplication with internal constraints becomes a recursive procedure, until no pairs $(r, l)$ are found and multiplication is performed by concatenation of each clique.

\textbf{Division of partially commutative words.} Divisibility of $v$ by $u$ is checked by finding a pair of reconstructible \texttt{clique\_words} $(l,r)$ such that $lur = v$. For this, we find all pairs $(l_i,r_i)$ of non-commutative words on $i$-th clique such that $l_i u_i r_i =v_i$ (this is a substring matching problem), where $u_i$ and $v_i$ are the projections of $u$ and $v$ to $i$-th clique and store them in a set $P_i$. Then for each possible pair of clique words $(l, r) \in \prod_i P_i$ with the above clique projections, we check if the clique words $l$ and $r$ are reconstructible.
This is done with the procedure mentioned in~\cite{liu1990efficient}.

\textbf{Tracial equivalence of partially commutative words}. \texttt{PCPOP} implements the algorithm from \cite{liu1990efficient} described in Section~\ref{sec:pcwords} to decide when two partially commutative words are equivalent up to cyclic permutations. The function \texttt{cyclic\_reduce} first reduces a given monomial in clique representation with respect to possibly additional constraints such as projectors, unipotents or unitaries. 
This reduction proceeds similar to the reduction in the multiplication procedure. It outputs the reduced clique representation, using data structure \texttt{CyclicWord}. This structure is used to compare the exponents of two partially commutative words and check if one divides a power of the other~\cite{liu1990efficient}.
\begin{example}
\label{ex:mult_pcmonomials}
Consider the alphabet $\vec x = \{a_0, a_1, b_0, b_1\}$ where $a_i$ commutes with $b_j$ and each letter is a projector, and monomials $u = a_0 a_1 b_0$ and $v = b_1 a_1 a_0$. Their product is the monomial $uv = a_0 a_1 a_0 b_0 b_1$. To obtain it from the clique normal forms, the multiplication algorithm looks at the cliques and the left and right edges of the monomials: 
\begin{align}
    u & = (a_0a_1, b_0) & u_l & = \{a_0, b_0\} & u_r & = \{a_1, b_0\}\, , \\
    v & = (a_1a_0, b_1) & v_l & = \{a_1, b_1\} & v_r & = \{a_0, b_1\} \, .
\end{align}
To multiply $u$ and $v$, we first multiply $u_r$ and $v_l$. This step also implements the special constraints (projections, unipotents and unitaries). For that, we check clique by clique if there is some combination of letters that gives rise to a constraints. In this case, this happens on the first clique: $a_1 a_1 = a_1$. We remove $a_1$ both from $u$ and $v$, call the results $u'$ and $v'$. Then we multiply $u'\mult a_1 \mult v'$. When there are no pairs in the right and left words corresponding to a constraint, we simply concatenate each clique of $u$ and $v$. This ensures that the recursive procedure for multiplication terminates.
\end{example} 

\section{Graph products}
\label{sec:graph_product_implementation}
Graph product monoids and graph product monomials have the dedicated data structures \texttt{GraphProductMonoid\{V\}} and \texttt{GraphProductWord\{V\}}, where $V$ is either \texttt{NCWord} or \texttt{GraphProductWord}. This recovers partially commutative monoids and monomials when $V$ is the type \texttt{Variable}. The data structure \texttt{PCMonomial} corresponds with mapping variables in \texttt{GraphProductWord} to unsigned integers, which makes arithmetical computation more efficient. As before,
\texttt{GraphProductWord\{V\}} contains three fields: the clique words \texttt{clique\_words} stored as an array of arrays, the initial letters \texttt{edge\_l}, and the final letters \texttt{edge\_r}. Multiplication is performed with the same algorithm discussed above for partially commutative words. The only difference is that now letters are replaced by monoid elements and concatenation of letters is replaced by multiplication of elements inside their corresponding monoid. 

\begin{example}
Consider three non-commutative monoids $\monoid_A = \langle a_0, b_0 \rangle$, $\monoid_B = \langle b_0, b_1\rangle$ and $\monoid_{AB} = \langle c_0, c_1\rangle$, where all the letters are projectors. Let $\monoid$ be their graph product with respect to the commutativity graph whose only edge is $(\monoid_A, \monoid_B)$. That is, the \emph{monoid cliques} $\{\monoid_A,\monoid_{AB}\}$ and $\{\monoid_B,\monoid_{AB}\}$ in the graph product monoid play the role of the maximal cliques in a partially commutative monoid. Take the words $u = c_0 b_0 a_1 a_0$ and $v = a_1 b_0 c_1$. The clique words, initial and final letters in the clique representation are: 
\begin{align}
    u & = ((c_0)(a_1a_0), (c_0)(b_0)) & u_l & = \{(c_0)\} & u_r & = \{(b_0), (a_1a_0)\} \, , \\
    v & = ((a_1)(c_1), (b_0)(c_1)) & v_l & = \{(a_1), (b_0)\} & v_r & = \{(c_1)\} \, .
\end{align}
Their product $u \mult v = (c_0)\mult (b_0)\mult (a_1 \mult a_0 \mult a_1)\mult (c_1)$, is obtained from the multiplication algorithm for clique words. First, this multiplies $u_r$ and $v_l$ checking clique by clique if the elements in right and left belong to the same monoid. If this is the case, we remove such elements from $u$ and $v$ to obtain $u'$ and $v'$ and multiply them inside their corresponding monoids to obtain $m$. In this example, $u' = ((c_0), (c_0))$, $v' = ((c_1), (c_1))$  and $m = ((a_1a_0a_1),(b_0))$. Finally, we multiply $u' \mult m \mult v'$. There are no more elements from the same monoids in $u'_r$ and $m_l$ nor $m_r$ and $v'_l$, therefore we simply concatenate the monoids in each clique to obtain $uv = ((c_0)(a_1a_0a_1)(c_1), (c_0)(b_0)(c_1))$.
\label{ex:multiplication_graph_product}
\end{example}

In a non-commutative monoid, multiplication is delegated to the multiplication algorithm in \texttt{AbstractAlgebra}, which automatically implements polynomial constraints via Gr\"obner bases computation algorithms. In partially commutative monoids, multiplication is implemented through the multiplication algorithm for \texttt{PCMonomial} described in Example~\ref{ex:mult_pcmonomials}. In a graph product monoid, the multiplication is implemented with the algorithm described in Example~\ref{ex:multiplication_graph_product}, where the multiplication inside each vertex monoid is delegated to its corresponding multiplication algorithm.

\section{Trace and state monoids}
\label{sec:tracemonoid}

Trace monoids and state monoids are special cases of partially commutative monoids, where letters under the (tracial) state symbol commute with everything and inherit the involution from the free generators. In order to make computations with trace and state monomials more convenient, these have a dedicated data structure \texttt{TraceMonoid}. Trace monoids are state monoids with an additional tracial condition, therefore we use the same structure with one Boolean field \texttt{tracial} that decides if the tracial condition is imposed. 
The fields \texttt{base\_monoid} and \texttt{state\_monoid} store the monoid of generators and the (tracial) state monoid as \texttt{GraphProductMonoid}. Additionally, \texttt{dict\_free} and \texttt{dict\_states} are dictionaries that map the variables in the base monoid to the (tracial) state monoid, and monomials in the base monoid to the variables in the (tracial) state monoid corresponding to that monomial under the (tracial) state symbol.

The structure \texttt{TraceMonoid} can be built from the base monoid including (tracial) state words up to a fixed degree using the function \texttt{make\_trace\_monoid}.
\texttt{PCPOP} provides some additional functionalities to ease computations with (tracial) state monoids. For instance, the function \texttt{state\_embedding} embeds polynomials in the base monoid to the (tracial) state monoid, and the function \texttt{state\_projection} puts the (tracial) state symbol on the free part of a (tracial) state word.
\chapter{Tutorial}
\label{ch:tutorial}

\section{Installation}
\label{sec:installation}

\texttt{PCPOP} can be readily installed with Julia in-built package manager:

\begin{juliacode}
\begin{lstlisting}[style=code, mathescape=true]
import Pkg
Pkg.add("PCPOP")
\end{lstlisting}
\end{juliacode}

After installation, it can be imported and immediately used:
\begin{juliacode}
\begin{lstlisting}[style=code, mathescape=true]
using PCPOP
\end{lstlisting}
\end{juliacode}

\section{List of functions}
\label{sec:functions}

\textbf{Building the polynomial algebras}.

\begin{enumerate}
    \item The macro \texttt{@ncmonoid M x[n, m]} initializes a non-commutative monoid \texttt{M} with $n$ hermitian and $m$ non-hermitian variables collected in \texttt{x}. Alternatively, the syntax \texttt{@ncmonoid M a[n,m] b[n,m] ...} initializes a non-commutative monoid with $n$ hermitian and $m$ non-hermitian variables collected in each \texttt{a, b, \ldots}.
    \item The function \texttt{add\_relations!(R)} internally sets the collection of constraints \texttt{R} in the base non-commutative monoid \texttt{M}, which are implemented at the level of arithmetical computations through Gr\"obner basis reductions computed with the package \texttt{AbstractAlgebra}.
    \item The macro \texttt{@pcmonoid M x[n, m]} initializes a \texttt{GraphProductMonoid} \texttt{M} with $n$ hermitian and $m$ non-hermitian variables collected in \texttt{x}.
    \item The function \texttt{change\_display(n)} changes how elements of \texttt{GraphProductMonoid} are displayed. When $n=1$ (default), the clique representation is displayed. When $n=2$, the clique representation is displayed but empty cliques are suppressed for the ease of readability. When $n=3$, a non-commutative representative of the equivalence class is displayed. Notice that the last display is obtained constructing the occurrence graph from the clique representation, therefore it is discouraged during intermediate computations.
    \item The macro \texttt{@comms a b} internally sets the commutation relation \texttt{a*b - b*a = 0}, which implements the constraint at the level of arithmetical computations inside the parent monoid \texttt{M}. When \texttt{a} and \texttt{b} are collections of variables, this sets relations between each letter in \texttt{a} and \texttt{b}. Alternatively, \texttt{@comms a b c \ldots} sets pairwise commutation relations.
    \item The macro \texttt{@ortho a b} internally sets the orthogonality relation \texttt{a*b = 0}, which implements the constraint at the level of arithmetical computations in the parent monoid. Notice that it does not automatically set \texttt{b*a = 0}.
    \item The function \texttt{adjoint(m)}, alternatively \texttt{m'}, gets the adjoint of the monomial \texttt{m}.
    \item The functions \texttt{Projector(x)}, \texttt{Unipotent(x)} and \texttt{Unitary(x)} set the relations \texttt{x*x - x = 0}, \texttt{x*x - 1 = 0} and \texttt{x*x' - 1 = 0} respectively, which implements the equality constraint at the level of arithmetical computations inside the parent monoid \texttt{M}.
    \item The function \texttt{GraphProductMonoid("M$\otimes$N", [M, N])} initializes a parent monoid \texttt{M$\otimes$N} that has all variables in \texttt{M} and \texttt{N} as variables. Commutation relations among the children monoids can be set with \texttt{@comms M N}.
    \item The function \texttt{build(M)} builds the monoid with the prescribed relations. It is necessary to build the monoid before performing arithmetical computations. Once the monoid is built, it is no longer possible to change it. This is to avoid possible issues: changing the structure of the monoid would require to update the representation of previously constructed elements. In the process of building a non-commutative monoid \texttt{M} with internal constraints, a (truncated) Gr\"obner basis for the constraints is computed and stored.
    \item The function \texttt{make\_trace\_monoid(M, d, tracial)} initializes the (tracial) state monoid with all (tracial) state monomials over the base monoid \texttt{M} up to degree \texttt{d}. The Boolean argument \texttt{tracial} decides if tracial conditions are imposed.
\end{enumerate}

A non-commutative monoid \texttt{M} has type \texttt{NCMonoid}. All variables in \texttt{M} are collected in the field \texttt{M.vertices}. All relations among the variables are collected in the field \texttt{M.relations}. A partially-commutative monoid \texttt{P} has type \texttt{GraphProductMonoid}. All variables in \texttt{P} are collected in the field \texttt{P.vertices}. Commutation relations are collected in the field \texttt{P.commutations}. A (tracial) state monoid \texttt{TM} has type \texttt{TraceMonoid}. The corresponding partially-commutative monoid is in the field \texttt{TM.state\_monoid}. All variables in the base monoid are collected in the field \texttt{TM.vertices\_free}, and all new variables corresponding to (tracial) state monomials are collected in the field \texttt{TM.vertices\_states}.

\begin{juliacode}
\begin{lstlisting}[style=code, mathescape=true]
# Initialize non-commutative monoid with 2 variables
@ncmonoid M x[2,0]
# Set variables to projectors
Projector.(x)
# Build the monoid
build(M)
# Arithmetical computations in M
p = x[1]*x[1] + x[2]*x[2]
\end{lstlisting}
\end{juliacode}

\begin{juliacode}
\begin{lstlisting}[style=code, mathescape=true]
# Initialize partially-commutative monoid with 4 variables
@pcmonoid M a[2,0] b[2,0]
# Set variables to unitaries
Unipotent.([a;b])
# Set commutation relations
@comms a b
# Build the monoid
build(M)
# Arithmetical computations in M
p = a[1]*b[1] + a[1]*b[2] + a[2]*b[1] - a[2]*b[2]
\end{lstlisting}
\end{juliacode}

\textbf{Building polynomial optimization relaxations.}

\begin{enumerate}
\label{list:pcpop}
    \item The function \texttt{pcpop(p, k)} builds the semidefinite programming relaxation of level \texttt{k} [Equation~\eqref{eq:sdp_relax}] for the polynomial optimization problem with objective function \texttt{p} inside its parent monoid \texttt{M}, which may internally carry relations among its variables. 
    The function returns 
    \begin{enumerate}[nosep, label=-]
        \item the objective value (when \texttt{optimize=true})
        \item a \texttt{JuMP.model}
        \item a dictionary that maps monomials to the scalar variable corresponding to its moment (an entry in the moment matrix when primal is true or the sum of all entries corresponding to that monomial when primal is false)
        \item the sequence of monomials indexing the localizing matrices
        \item the sequence of monomials indexing the principal moment matrix
    \end{enumerate}
    Alternatively, the method \texttt{pcpop(p, basis, basis\_principal)} admits a custom indexing set for the moment matrix \texttt{basis\_principal} and a custom indexing set for the localized moment matrices \texttt{basis}. Both methods admit the following additional keyword arguments:
    \begin{enumerate}[nosep, label=-]
        \item \texttt{min}: Boolean value for the optimization sense (default \texttt{false} - maximizes).
        \item \texttt{op\_eq}: list of additional polynomial equality constraints not internally imposed in the monoid (default empty vector).
        \item \texttt{op\_ge}: list of polynomial inequality constraints (default empty vector).
        \item \texttt{tr\_eq}: list of linear equality constraints on the moments. It takes a vector of tuples, where each tuple $(p,v)$ contains a polynomial $p$ and a scalar value $c$ corresponding with the moment constraint $L(p)=c$ (default empty vector).
        \item \texttt{tr\_ge}: list of linear inequality constraints on the moments. Same format as \texttt{tr\_eq}, each tuple $(p,c)$ contains a polynomial $p$ and a scalar value $c$ corresponding with the moment constraint $L(p) \geq c$ (default empty vector).
        \item \texttt{lvl\_lm}: integer value for the level of the localizing matrices. The default value $-1$ builds the localizing matrices using the monomial basis of degree \texttt{k} and extends the monomial basis for the moment matrix in order for it to contain all monomials appearing in the localizing matrices. A fixed value $d $ builds the localizing matrices using the monomial basis of degree $d$ and throws an error if the localizing matrices contain any monomial not appearing in the moment matrix.
        \item \texttt{list\_vars}: list of variables used to compute the monomial bases for the moment matrices. By default, it is empty and all the variables appearing in  \texttt{p}, \texttt{op\_eq}, \texttt{op\_ge}, \texttt{tr\_eq} and \texttt{tr\_ge} are used.
        \item \texttt{tracial}: Boolean value to impose cyclic equivalence (default \texttt{false}).
        \item \texttt{normalize}: Boolean value to impose the normalization $L(\id) = 1$ (default \texttt{true}).
        \item \texttt{solver}: sets a solver \texttt{JuMP.Optimizer}, (default \texttt{Mosek.Optimizer}).
        \item \texttt{model\_flags}: array of flags for the model (default empty).
        \item \texttt{optimize}: Boolean value to optimize the model (default \texttt{true}).
        \item \texttt{reduce}: Boolean value to implement Jordan algebra reduction (default \texttt{false}).
        \item \texttt{block\_diag}: Boolean value to implement a numerical block-diagonalization of the semidefinite program (default \texttt{false}).
        \item \texttt{primal}: Boolean value to decide the primal (moment relaxation) or dual (sum of squares relaxation) implementation (default \texttt{true}).
        \item \texttt{canonical}: Boolean value deciding the implementation of the semidefinite relaxation. When \texttt{true} (default) it implements the matrix variable version in Equation~\eqref{eq:sdp_matrix}, when \texttt{false} it implements the scalar variable version in Equation~\eqref{eq:sdp_scalar}.
    \end{enumerate}
    The primal program implements the moment relaxation in Equation~\eqref{eq:sdp_relax}. The dual program implements the sum of squares relaxation in Equation~\eqref{eq:sdp_sos_relax}. Although equivalent, the performance and numerical stability can depend on the implementation. This will be apparent in the benchmarking example~\ref{tab:benchmark_polygon_bell}. The block-diagonalization uses the numerical algorithm based on randomization \cite[Algorithm 4.1]{murota2007numerical} implemented in \texttt{SDPSymmetryReduction}. Although when successful this significantly reduces the cost of solving the semidefinite program, it may fail due to rounding errors. 
    \item The method \texttt{pcpop(p, k, G, action)} admits as additional arguments a group \texttt{G} of symmetries and its action \texttt{action} over \texttt{M}. It builds the symmetry reduced sum of squares relaxation of level \texttt{k} for the polynomial optimization problem with objective function \texttt{p} using \texttt{SymbolicWedderburn} and returns a \texttt{JuMP.model}. The group action must have abstract type \texttt{SymbolicWedderburn.Action}.
    \item The function \texttt{tpop(p, TM, basis)} builds the sum of squares relaxation with monomial basis \texttt{basis} for the (tracial) state polynomial optimization problem with objective function \texttt{p} in the (tracial) state monoid \texttt{TM}. The keyword arguments are similar as those for \texttt{pcpop}. 
\end{enumerate}

Examples of each of these methods are discussed in the following sections. 

\section{Non-commutative polynomial optimization}
\label{sec:ncpop}

Consider the following non-commutative polynomial optimization problem:
\begin{align}
    \sup \quad & a_1 b_1 + a_1 b_2 + a_2 b_1 - a_2 b_2 \label{eq:ncpop_chsh} \\
    \operatorname{s.t.} \quad & a_0^2 = 1 & a_0 b_0 = b_0 a_0 \, , & & & & \nonumber \\
    & a_1^2 = 1 & a_1 b_0 = b_0 a_1 \, , & & & &\nonumber \\
    & b_0^2 = 1 & a_0 b_1 = b_1 a_0 \, , & & & & \nonumber \\
    & b_1^2 = 1 & a_1 b_1 = b_1 a_1 \, . & & & & \nonumber
\end{align}

The first level relaxation is indexed with the degree one monomials $(\id, a_0, a_1, b_0, b_1)$:
\begin{align}
    \sup \quad & L(a_1 b_1) + L(a_1 b_2) + L(a_2 b_1) - L(a_2 b_2) \label{eq:chsh_level_1} \\
    \operatorname{s.t.} \quad &  
    \begin{pmatrix}
    1 & * & *  & *  & * \\
      & 1 & * & L(a_0b_0) & L(a_0b_1) \\
      &   & 1 & L(a_1b_0) & L(a_1b_1) \\
      &   &   & 1 & * \\
      &   &   &   & 1 
    \end{pmatrix} \geq 0
    \, . \nonumber
\end{align}
The optimal value of the semidefinite relaxation in Equation~\eqref{eq:chsh_level_1} is $2.8284...$, which gives an upper bound to the optimal value of Problem~\eqref{eq:ncpop_chsh}. In this case, the bound is already tight up to numerical precision. The implementation in \texttt{PCPOP} is shown below. We build a \texttt{NCMonoid} that internally carries all polynomial constraints, and use the function \texttt{pcpop} to build and solve the semidefinite programming relaxation.

\begin{juliacode}
\begin{lstlisting}[style=code, mathescape=true]
# Initialize non-commutative monoid with 4 variables
@ncmonoid M a[2,0] b[2,0]
# Internal equality constraints
R = [a[1]^2 - 1,
     a[2]^2 - 1,
     b[1]^2 - 1,
     b[2]^2 - 1,
     a[1]*b[1] - b[1]*a[1],
     a[1]*b[2] - b[2]*a[1],
     a[2]*b[1] - b[1]*a[2],
     a[2]*b[2] - b[2]*a[2]]
add_relations!(R)
# Build monoid
build(M)
# Objective function
p = a[1]*b[1] + a[1]*b[2] + a[2]*b[1] - a[2]*b[2]
# Optimization of the semidefinite relaxation
val, model, _ = pcpop(p, 1) 
println("Optimal value is ", val)
\end{lstlisting}
\end{juliacode}

\section{Partially commutative polynomial optimization}
\label{sec:pcpop}

We now consider the non-commutative polynomial optimization problem in Equation~\eqref{eq:ncpop_chsh} as a partially-commutative polynomial optimization problem. In this scenario, the commutation relations and the unipotent constraints are internally implemented at the level of arithmetical computations within the partially-commutative monoid instead of through reductions. The implementation in \texttt{PCPOP} is shown below. 
We build a \texttt{GraphProductMonoid} that internally carries the commutation relations and unipotency constraints, and use the function \texttt{pcpop} to build and solve the semidefinite relaxation.

\begin{juliacode}
\begin{lstlisting}[style=code, mathescape=true]
# Initialize partially-commutative monoid with 4 variables
@pcmonoid M a[2,0] b[2,0]
# Set variables to unitaries
Unipotent.([a;b])
# Set commutation relations
@comms a b
# Build the monoid
build(M)
# Objective function
p = a[1]*b[1] + a[1]*b[2] + a[2]*b[1] - a[2]*b[2]
# Optimization of the semidefinite relaxation
val, model, _ = pcpop(p, 1) 
println("Optimal value is ", val)
\end{lstlisting}
\end{juliacode}


\section{Commutative polynomial optimization}
\label{sec:tutorial_pop}

Commutative polynomial optimization problems can be recovered as a special case of partially-commutative polynomial optimization problems, in which all variables commute. In this scenario, the dependence graph has no edges and each vertex forms a maximal clique. Therefore, our clique or graph representations simply count the number of occurrences of each letter, which is the standard exponent representation for commutative monomials. Consider the following commutative polynomial optimization problem from \cite[Example 5]{lasserre2001global}.
\begin{align}
    \inf \quad & - (a - 1)^2 - (a - b)^2 - (b - 3)^2
\label{eq:pop_tutorial} \\
    \operatorname{s.t.} \quad & 1-(a - 1)^2 \geq 0 \, , \nonumber \\
    & 1-(a - b)^2 \geq 0 \, , \nonumber \\
    & 1-(b - 3)^2 \geq 0 \, . \nonumber
\end{align}
The optimal value $-2$ is attained with the second level semidefinite relaxation. The implementation in \texttt{PCPOP} is shown below (discarding the constant term in the objective). We build a \texttt{GraphProductMonoid} that internally carries the commutation relations among all variables, recovering the exponent representation of commutative monomials. We explicitly provide the polynomial inequalities to the function \texttt{pcpop}, which builds and solves the semidefinite relaxation.

\begin{juliacode}
\begin{lstlisting}[style=code, mathescape=true]
# Build commutative monoid with 2 variables
@pcmonoid M a b
@comms a b
build(M)
# Objective function
p = - (a - 1)^2 - (a - b)^2 - (b - 3)^2 + 10
# Inequality constraints
S = [1 - (a - 1)^2, 1 - (a - b)^2, 1 - (b - 3)^2]
# Optimization of the semidefinite relaxation
val, model, _ = pcpop(p, 2; op_ge=S,min=true)
println("Optimal value is   ", val)

\end{lstlisting}
\end{juliacode}

\section{Tracial polynomial optimization}
\label{sec:tutorial_tpop}

Consider the following example of tracial polynomial optimization \cite[Example 5.14]{burgdorf2016optimization}. 
\begin{align}
    \inf \quad & (1-a^2)(1-b^2) + (1-b^2)(1-a^2)
\label{eq:tracial_pop_tutorial} \\
    \operatorname{s.t.} \quad & \id - a^2 \geq 0 \, , \nonumber \\
    & \id - b^2 \geq 0 \, . \nonumber
\end{align}
The optimal value is $0$. Third level semidefinite relaxation gives a lower bound $-0.0031$, and fourth level semidefinite relaxation a lower bound $-0.0010$. These bounds match the ones in \cite{burgdorf2016optimization} with one extra level. This discrepancy is due to the implementation. By default, \texttt{PCPOP} admits extra monomials in the relaxation appearing in the localizing matrices for the constraints, providing bigger and tighter relaxations. The implementation in \texttt{PCPOP} is shown below. We build a \texttt{GraphProductMonoid} with no internal relations. We use \texttt{pcpop} to build and solve the semidefinite relaxation, setting the tracial optimization with the keyword argument \texttt{tracial=true}.
\begin{juliacode}
\begin{lstlisting}[style=code, mathescape=true]
# Build partially-commutative monoid with 2 variables
@pcmonoid M a b
build(M)
# Objective function
p = (1 - a^2)*(1 - b^2) + (1 - b^2)*(1 - a^2)
# Inequality constraints
S = [1 - a^2, 1 - b^2]
# Optimization of the semidefinite relaxation
val, model, _ = pcpop(p, 3, op_ge=S, tracial=true,min=true)
println("Optimal value is ", val)
\end{lstlisting}
\end{juliacode}

\section{State polynomial optimization}
\label{sec:tutorial_spop}

Consider the example of state polynomial optimization from \cite[Example 7.2.1]{klep2024state}, which corresponds with the quadratic Bell inequality proposed in \cite{uffink2002quadratic}. Namely,
\begin{align}
    \sup \quad & (\rho(a_1 b_2) + \rho(a_2 b_1))^2 + (\rho(a_1 b_1) - \rho(a_2 b_2))^2 
\label{eq:spop_tutorial} \\
    \operatorname{s.t.} \quad & a_0^2 = 1 \, , \hspace{3em} a_0 b_0 = b_0 a_0 \, , \nonumber \\
    & a_1^2 = 1 \, , \hspace{3em} a_1 b_0 = b_0 a_1 \, , \nonumber \\
    & b_0^2 = 1 \, , \hspace{3em}  a_0 b_1 = b_1 a_0 \, , \nonumber \\
    & b_1^2 = 1 \, , \hspace{3em} a_1 b_1 = b_1 a_1 \, . \nonumber
\end{align}
We obtain the optimal value $4$ at level three relaxation up to numerical precision. The implementation in \texttt{PCPOP} is shown below. 
We build a \texttt{GraphProductMonoid} for the free generators that carries the commutations and unipotent constraints. In order to encode the state polynomial in the objective function, we build a \texttt{TraceMonoid} over the base generators using \texttt{make\_trace\_monoid}. Then, we use \texttt{tpop} to build the semidefinite relaxation.

\begin{juliacode}
\begin{lstlisting}[style=code, mathescape=true]
using JuMP,Mosek,MosekTools
# Initialize partially-commutative monoid with 4 variables
@pcmonoid M a[2,0] b[2,0]
# Set variables to projectors
Unipotent.(a)
Unipotent.(b)
@comms a b
# Build the monoid
build(M)
# Build new monoid with state monomials up to degree 6
TM = make_trace_monoid(M, 6, tracial=false)
# Objective function
p  = (state(a[1]*b[2], TM) + state(a[2]*b[1], TM))^2 
p += (state(a[1]*b[1], TM) - state(a[2]*b[2], TM))^2
# Basis for the semidefinite relaxation
basis = trace_monomials(TM, 0:3)
# Build sum of squares relaxation
sos_model = tpop(p, TM, basis)
# Optimization of the semidefinite relaxation
set_optimizer(sos_model, Mosek.Optimizer)
optimize!(sos_model)
println("Optimal value is ", objective_value(sos_model))
\end{lstlisting}
\end{juliacode}

\section{Trace polynomial optimization}
\label{sec:tutorial_tracepop}

Consider the example of trace polynomial optimization from \cite[Example 6.1]{klep2022optimization}.
\begin{align}
    \inf \quad & \rho(abc) + \rho(ab)\rho(c)
\label{eq:tpop_tutorial} \\
    \operatorname{s.t.} \quad & a^2 = a \, , \nonumber \\
    & b^2 = b \, , \nonumber \\
    & c^2 = c \, . \nonumber
\end{align}
The optimal value $-1/32$ is attained at the level three semidefinite relaxation up to numerical precision. The implementation in \texttt{PCPOP} is shown below. 
As before, we build a \texttt{TraceMonoid} over a \texttt{GraphProductMonoid} of free generators and use \texttt{tpop} to build the semidefinite relaxation. However, we now set the keyword argument \texttt{tracial=true} to enforce the tracial equivalences both in the trace polynomial monoid and the polynomial optimization problem.

\begin{juliacode}
\begin{lstlisting}[style=code, mathescape=true]
# Initialize partially-commutative monoid with 3 variables
using JuMP,Mosek,MosekTools
@pcmonoid M a b c
# Set variables to projectors
Projector(a)
Projector(b)
Projector(c)
# Build the monoid
build(M)
# Build new monoid with trace monomials up to degree 6
TM = make_trace_monoid(M, 6, tracial=true)
# Objective function
p = - state(a*b*c, TM) - state(a*b, TM)*state(c, TM)
# Basis for the semidefinite relaxation
basis = trace_monomials(TM, 0:3, tracial=true)
# Build sum of squares relaxation
model = tpop(p, TM, basis, tracial=true)
# Optimization of the semidefinite relaxation
set_optimizer(model, Mosek.Optimizer)
optimize!(model)
println("Optimal value is ", objective_value(model))
\end{lstlisting}
\end{juliacode}

\section{Symmetry reductions}
\label{sec:tutorial_symmetries}

As we discussed in Section~\ref{sec:symmetries}, when a polynomial optimization problem has symmetries, it is possible to reduce to problem to an invariant subspace. Consider, for instance, Problem~\eqref{eq:ncpop_chsh} corresponding to the quantum value of CHSH inequality
\begin{align}
    \sup \quad & a_1 b_1 + a_1 b_2 + a_2 b_1 - a_2 b_2 \\
    \operatorname{s.t.} \quad & a_0^2 = 1 & a_0 b_0 = b_0 a_0 \, , & & & & \nonumber \\
    & a_1^2 = 1 & a_1 b_0 = b_0 a_1 \, , & & & &\nonumber \\
    & b_0^2 = 1 & a_0 b_1 = b_1 a_0 \, , & & & & \nonumber \\
    & b_1^2 = 1 & a_1 b_1 = b_1 a_1 \, . & & & & \nonumber
\end{align}
This problem is invariant under the exchange of the two parties (among other symmetries). Therefore, it is invariant under the group $G = \langle \id, \pi\rangle$ with transformations
\begin{align}
    & \id :(a_0, a_1, b_0, b_1) \to (a_0, a_1, b_0, b_1) \, . \\
    & \pi :(a_0, a_1, b_0, b_1) \to (b_0, b_1, a_0, a_1) \, . \nonumber
\end{align}
The symmetrized first level relaxation has $16$ variables and $9$ constraints, while the non-symmetrized relaxation in Equation~\eqref{eq:chsh_level_1} has $16$ variables and $14$ constraints. The implementation in \texttt{PCPOP} is shown below. 
The group actions have abstract type \texttt{Action} required for the symmetrization with \texttt{SymbolicWedderburn}. Although alternative actions to encode more general polynomial automorphisms can be defined, for most applications we use the action \texttt{OnLetters} that encodes how a permutation on the alphabet acts over monomials and polynomials. We use \texttt{PermutationGroups} to encode permutations.

\begin{juliacode}
\begin{lstlisting}[style=code, mathescape=true]
using JuMP, Mosek, MosekTools
import PermutationGroups as PG
# Build monoid
@pcmonoid M a b c d
@comms [a, b] [c, d]
Unipotent.([a,b,c,d])
build(M)
# Group action
action = OnLetters()
π = PG.perm"(1,3)(2,4)"
G = PG.PermGroup(π)
# Optimize symmetrized semidefinite relaxation
p = a*c + a*d + b*c - b*d
model = pcpop(p, 1, G, action)
set_optimizer(model, Mosek.Optimizer)
set_silent(model)
optimize!(model);
println("Objective value is  ", objective_value(model))
\end{lstlisting}
\end{juliacode}

The symmetry reduction considers the constraints internally set in the monoid, which include commutations, projections, unitaries, unipotents and orthogonalities. Although symmetry reductions can be extended to polynomial optimization problems with additional constraints without significant complications, these are not currently implemented.

\section{Jordan algebra reductions}
\label{sec:jordan}

Finding the symmetries of a polynomial optimization problem may be a challenging problem in itself. Jordan algebra reductions provide alternative axiomatic reductions without explicitly considering the symmetries \cite{permenter2020dimension, brosch2022jordan}. As an example, consider again the polynomial optimization problem corresponding with the maximal quantum value of CHSH functional discussed in Sections~\ref{sec:tutorial_pop} and~\ref{sec:tutorial_symmetries}. The first level semidefinite relaxation in Equation~\eqref{eq:ncpop_chsh} involves $13$ different monomials. Jordan algebra reduction provides an invariant subspace spanned by $7$ variables without explicitly specifying any symmetries.
The implementation in \texttt{PCPOP} is shown below. 
We build a \texttt{GraphProductMonoid} that internally carries the commutation and unipotent constraints, and we use the function \texttt{pcpop} to build and solve the reduced semidefinite relaxation.
The keyword argument \texttt{reduce=true} formats the semidefinite program in vectorized canonical form and uses \texttt{SDPSymmetryReduction} to obtain an invariant subspace, and the keyword argument  \texttt{block\_diag=true} block-diagonalizes the invariant subspace.

\begin{juliacode}
\begin{lstlisting}[style=code, mathescape=true]
# Build monoid
@pcmonoid M a b c d
@comms [a, b] [c, d]
Unipotent.([a,b,c,d])
build(M)
# Objective function
p = a*c + a*d + b*c - b*d
# Jordan algebra reduction
val, model, _ = pcpop(p, 1; reduce=true)
println("Objective value is  ", objective_value(model))
\end{lstlisting}
\end{juliacode}
The reduced semidefinite program involves $7$ scalar variables and one positive semidefinite constraint of size $5$. The block-diagonalized problem involves instead two positive semidefinite constraints of size $2$, and one scalar constraint.
\begin{juliacode}
\begin{lstlisting}[style=code, mathescape=true]
# Block diagonalization of Jordan reduction
val, _ = pcpop(p, 1; reduce=true, block_diag=true)
println("Objective value is  ", val)
\end{lstlisting}
\end{juliacode}

Although the first level is already tight, for the purpose of illustration let us consider the second level relaxation, which is spanned by $13$ monomials. The primal implementation has $91$ variables, $31$ linear constraints and one semidefinitze constraint of size $13$. The Jordan reduction has $97$ variables, $61$ linear constraints and one semidefinite constraint of size $13$. Therefore, in this case the invariant subspace obtained exploiting the algebraic structure in the polynomial algebra from the beginning is better than the one obtained with the Jordan reduction.
The numerical block-diagonalization of the invariant subspace requires complex matrices, and produces two semidefinite constraints of size $9$ and $4$ instead.
\begin{juliacode}
\begin{lstlisting}[style=code, mathescape=true]
# Block diagonalization of Jordan reduction
val, _ = pcpop(p, 2; reduce=true,
                     block_diag=true,
                     complex=true)
println("Objective value is  ", val)
\end{lstlisting}
\end{juliacode}
\chapter{Applications}
\label{ch:applications}

In this chapter we review a collection of polynomial optimization problems that appear in current research in quantum information science. These include: characterizing quantum correlations in variations of Bell scenarios, contextuality scenarios and quantum networks, computing uncertainty relations and quantum relative entropies, and the security analysis of quantum cryptography protocols.

\section{Bell scenarios}
\label{sec:bell}

Bell scenarios are measurement protocols in which different parties perform local measurements over a shared physical state. Under certain independence assumptions, correlations in these scenarios can discriminate classical, quantum and post-quantum theories. Non-classical correlations offer advantages in different information theoretical tasks, which are exploited in a wide range of applications. Despite its fundamental and practical interest, it is generally hard to describe the sets of correlations compatible with quantum theory. Polynomial optimization provides effective approximate descriptions for these sets in different measurement protocols.

\subsection{Bell inequalities}

Consider the Bell scenario with two parties, each with access to two dichotomic measurements $a_0$, $a_1$ and $b_0, b_1$. The maximal quantum value of the Clauser-Horne-Shimony-Holt (CHSH) \cite{clauser1969proposed} functional is the optimal value of the polynomial optimzation problem
\begin{align}
    \sup \quad & a_0 b_0 + a_0 b_1 + a_1 b_0 - a_1 b_1 \label{eq:chsh} \\
    \operatorname{s.t.} \quad & a_i^2 = 1 & &\forall i \in \{0,1\} \nonumber\\ 
    & b_i^2 = 1 & &i \in \{0,1\}\nonumber \\
    & a_i b_j = b_j a_i & &\forall i,j \in \{0,1\} \nonumber
\end{align}

This is precisely the example that we considered in the previous chapter. The first level relaxation already achieves the optimal value $2\sqrt{2}$ up to numerical precision.
\begin{juliacode}
\begin{lstlisting}[style=code, mathescape=true]
# Initialize partially-commutative monoid with 4 variables
@pcmonoid M a[2,0] b[2,0]
# Set variables to unitaries
Unipotent.([a;b])
# Set commutation relations
@comms a b
# Build the monoid
build(M)
# Objective function
p = a[1]*b[1] + a[1]*b[2] + a[2]*b[1] - a[2]*b[2]
# Optimization of the semidefinite relaxation
val, model, _ = pcpop(p, 1) 
println("Optimal value is ", val)
\end{lstlisting}
\end{juliacode}

\subsection{Routed Bell Scenario}
Consider the modified Bell scenario introduced in \cite{chaturvedi2024extending, Lobo_2024}, where Bob possesses two measurement setups: a \emph{short range} setup situated close to the source of quantum states and a \emph{long range} setup situated far away. This routed configuration is designed to certify loophole-free Bell non-locality over long distances. Although it is experimentally challenging to keep quantum effects over large distances, strong quantum correlations between Alice and Bob's short range device (e.g. certified by a large CHSH value) can be exploited to put additional constraints over the classical correlations between Alice and Bob's long range device, which are termed \emph{short-range quantum} correlations.

As an example, consider the routed Bell scenario with two dichotomic measurements. The maximal short-range quantum value for CHSH between Alice and Bob's long range device conditioned to a maximal quantum value for CHSH between Alice and Bob's short range device corresponds with the optimal value of the following optimization problem~\cite{Lobo_2024} 
        \begin{align}
    \sup \quad & a_0 b_{0,L} + a_0 b_{1,L} + a_1 b_{0,L} - a_1 b_{1,L} \\
    \operatorname{s.t.} \quad & a_i^2 = \id & &\forall i \in \{0,1\} \nonumber\\ 
    & b_{i,S}^2 = \id & &i \in \{0,1\}\nonumber \\
    & b_{i,L}^2 = \id & &i \in \{0,1\}\nonumber \\
    & a_i b_{j,S} = b_{j,S} a_i & &\forall i,j \in \{0,1\} \nonumber\\
    & a_i b_{j,L} = b_{j,L} a_i & &\forall i,j \in \{0,1\} \nonumber \\
    & b_{i,L} b_{j,L} = b_{j,L} b_{i,L} & &\forall i,j \in \{0,1\} \nonumber \\
    & a_0 b_{0,S} + a_0 b_{1,S} + a_1 b_{0,S} - a_1 b_{1,S} = 2\sqrt{2} \nonumber
\end{align}
This has optimal value $2$, which is attained with the second level semidefinite relaxation up to numerical precision. The implementation in \texttt{PCPOP} is shown below. 
\begin{juliacode}
\begin{lstlisting}[style=code, mathescape=true]
# Initialize monoid with Alice and Bob (short and large) setups
@pcmonoid M A[2,0] BS[2,0] BL[2,0]
Unipotent.(M.vertices)
@comms A BS
@comms A BL
@comms BL # joint measurability
build(M)
# Objective function
obj = A[1]*BL[1] + A[1]*BL[2] + A[2]*BL[1] - A[2]*BL[2]
# Constraints
T = A[1]*BS[1] + A[1]*BS[2] + A[2]*BS[1] - A[2]*BS[2]
tr_eq = [[T, 2*sqrt(2)]]
# Semidefinite relaxation
val, model, _=pcpop(obj, 2; tr_eq=tr_eq)
println("Optimal value is ", val)
\end{lstlisting}
\end{juliacode}

\subsection{Genuine multipartite nonlocality}

    Consider the problem of certifying genuine multipartite nonlocality in a Bell scenario with $n$ parties and two dichotomic measurements proposed in \cite{adhikary2024self}. This problem involves only linear constraints on the expectation values of the operators under the state, therefore it is not necessary to consider the state polynomial optimization framework. The problem for $n = 2$ reads:
   \begin{align}
       \sup \quad & a_1 b_1 \label{ex:multipartite_nonlocality}\\
    \operatorname{s.t} \quad & [a_i,b_j] =0 & \forall i,j \in \{1,2\} \, , \nonumber \\
    & a_i^2-a_i = b_j^2-b_j = 0 & \forall i,j \in \{1,2\} \, , \nonumber \\
    & \rho(a_2 \mult b_1) = \rho(a_1 \mult b_2) = 0 \, , \nonumber \\
    & \rho((1-a_2) \mult (1-b_2)) = 0 \, . \nonumber
\end{align}
The second level relaxation matches the upper bound $0.0902$ obtained in \cite{adhikary2024self} up to numerical precision. The implementation in \texttt{PCPOP} is shown below.

\begin{juliacode}
\begin{lstlisting}[style=code, mathescape=true]
# Initialize partially-commutative monoid with 4 variables
@pcmonoid M a[2,0] b[2,0]
# Set variables to projectors
Projector.([a;b])
# Set commutation relations
@comms a b
# Build the monoid
build(M)
# Objective function
obj = a[1]*b[1]
# Linear constraints on the moments
tr_eq = [[a[2]*b[1], 0],
         [a[1]*b[2], 0],
         [(1-a[2])*(1-b[2]), 0]]
# Optimization of the semidefinite relaxation
val, model,_ = pcpop(obj, 2; tr_eq=tr_eq)
println("Optimal value is ", val)
\end{lstlisting}
\end{juliacode}
    
\subsection{Overlapping Bell scenario} 

Consider a physical system with two unitary operators acting over each of the three separate components $A$, $B$ and $C$, and two more unitary operators acting jointly over components $BC$. The maximal quantum value of three CHSH functionals among $A$ and $B$, $C$ and $BC$ is the optimal value of the polynomial optimization problem
  \begin{align}
       \sup \quad &  a_0(b_0 + b_1 + c_0 + c_1 + x_0 + x_1) + a_1(b_0 - b_1 + c_0 - c_1 + x_0 - x_1) \\
    \operatorname{s.t} \quad & [a_i,b_j] =0 \, \hspace{3em} a_i^2 = \id \, , \nonumber \\
    & [a_i,c_j] =0 \hspace{3em} b_i^2 = \id \,  , \nonumber \\
    & [b_i,c_j] =0 \hspace{3em} c_i^2 = \id \,  , \nonumber \\
    & [a_i,x_j] =0 \hspace{3em} x_i^2 = \id \,  . \nonumber
\end{align}
The effects of overlapping measurements in Bell scenarios can be used to witness physical dimensions \cite{moran2023bell}, but here we consider no constraints on the dimension.  The second level semidefinite relaxations already shows a form of Bell monogamy for these correlations: the optimal value is $2\sqrt{2} + 4$ up to numerical precision, which corresponds with one inequality attaining the maximal quantum value and the other two classical values. The implementation in \texttt{PCPOP} is shown below. The implementation with graph products offers a short-cut to encode commutation relations. Although there is no significant gain in this simple scenario, it is shown for illustrative purposes. 
\begin{juliacode}
\begin{lstlisting}[style=code, mathescape=true]
# Initialize local monoids
@ncmonoid A a1 a2
@ncmonoid B b1 b2
@ncmonoid C c1 c2
@ncmonoid BC x1 x2
Unipotent.([a1, a2, b1, b2, c1, c2, x1, x2])
# Build global monoid
M = GraphProductMonoid("M",[A, B, C, BC])
@comms A B C
@comms A BC
build(M)
# Objective function.
p  = a1*(b1 + b2) + a2*(b1 - b2)
p += a1*(c1 + c2) + a2*(c1 - c2)
p += a1*(x1 + x2) + a2*(x1 - x2) 
# Optimize semidefinite relaxation
val,_ = pcpop(p,2) 
println("Optimal value is   ", val)
\end{lstlisting}
\end{juliacode}

\subsection{Non-linear Bell inequalities} 

The maximal quantum value of a non-linear Bell inequality corresponds with the optimal value of a state polynomial optimization problem. As an illustration, we have already shown in Section~\ref{sec:spop} how to implement the state polynomial optimization problem corresponding with quadratic Bell inequality proposed in \cite{uffink2002quadratic}. Namely,
\begin{align}
    \sup \quad & (\rho(a_1 b_2) + \rho(a_2 b_1))^2 + (\rho(a_1 b_1) - \rho(a_2 b_2))^2 
\label{eq:spop_example} \\
    \operatorname{s.t.} \quad & a_0^2 = 1 \, , \hspace{3em} a_0 b_0 = b_0 a_0 \, , \nonumber \\
    & a_1^2 = 1 \, , \hspace{3em} a_1 b_0 = b_0 a_1 \, , \nonumber \\
    & b_0^2 = 1 \, , \hspace{3em}  a_0 b_1 = b_1 a_0 \, , \nonumber \\
    & b_1^2 = 1 \, , \hspace{3em} a_1 b_1 = b_1 a_1 \, . \nonumber
\end{align}
We obtain the optimal value $4$ at level three relaxation up to numerical precision. The implementation in \texttt{PCPOP} is again shown below.

\begin{juliacode}
\begin{lstlisting}[style=code, mathescape=true]
using JuMP,Mosek,MosekTools
# Initialize partially-commutative monoid with 4 variables
@pcmonoid M a[2,0] b[2,0]
# Set variables to projectors
Unipotent.(a)
Unipotent.(b)
@comms a b
# Build the monoid
build(M)
# Build new monoid with state monomials up to degree 6
TM = make_trace_monoid(M, 6, tracial=false)
# Objective function
p  = (state(a[1]*b[2], TM) + state(a[2]*b[1], TM))^2 
p += (state(a[1]*b[1], TM) - state(a[2]*b[2], TM))^2
# Basis for the semidefinite relaxation
basis = trace_monomials(TM, 0:3)
# Build sum of squares relaxation
sos_model = tpop(p, TM, basis)
# Optimization of the semidefinite relaxation
set_optimizer(sos_model, Mosek.Optimizer)
optimize!(sos_model)
println("Optimal value is ", objective_value(sos_model))
\end{lstlisting}
\end{juliacode}

\section{Contextuality scenarios}
\label{sec:contextuality}

Contextuality scenarios can be thought of as measurement protocols without subsystems, where instead of locality constraints there are some consistency conditions among the measurements. Correlations in contextuality scenarios also allow under certain assumptions to discriminate classical, quantum and post-quantum theories, and find several practical applications. Again, polynomial optimization techniques provide effective approximate descriptions for the sets of quantum correlations in contextuality scenarios.

\subsection{Magic square game}

As an example of a contextuality scenario, consider the magic square game \cite{peres1990incompatible, mermin1990simple}. The game asks whether there exist $9$ unitary operators assembled in a $3 \times 3$ matrix such that $(i)$ elements in each row commute and their product is the identity, and $(ii)$ elements in each column commute and their product is minus the identity. This can be posed as a polynomial feasibility problem:

\begin{align}
    \sup \quad & 0
\label{eq:tpop_example} \\
    \operatorname{s.t.} \quad & x_{ij}x_{ij}^* = \id \, , \nonumber \\
    & x_{ij}x_{ik} = x_{ik} x_{ij} \, , \nonumber \\
    & x_{ij}x_{kj} = x_{kj} x_{ij} \, , \nonumber \\
    & x_{i1}x_{i2}x_{i3} = \id \, , \nonumber \\
    & x_{1i}x_{2i}x_{3i} = -\id \, . \nonumber
\end{align}
The following collection of two qubit Pauli operators provides a feasible solution
\begin{equation}
\begin{array}{rrr}
    \phantom{-}I \otimes Z &  \phantom{-}Z \otimes I &  Z \otimes Z \\
    \phantom{-}X \otimes I & I \otimes X & \phantom{-}X \otimes X \\
    -X \otimes Z & -Z\otimes X & Y \otimes Y
\end{array} \, .
\end{equation}
Therefore, every semidefinite relaxation of Problem~\eqref{eq:tpop_example} must be feasible. The implementation in \texttt{PCPOP}  is shown below.
\begin{juliacode}
\begin{lstlisting}[style=code, mathescape=true]
# Build the monoid
using JuMP
@pcmonoid M X[9,0]
Unipotent.(X)
x = reshape(X,(3,3))
for i in 1:3
	@comms x[i, 1] x[i, 2] x[i, 3]
	@comms x[1, i] x[2, i] x[3, i]
end
build(M)
# Conditions on the game
R = [x[1,1]*x[1,2]*x[1,3] - 1,
	 x[2,1]*x[2,2]*x[2,3] - 1,
	 x[3,1]*x[3,2]*x[3,3] - 1,
	 x[1,1]*x[2,1]*x[3,1] + 1,
	 x[1,2]*x[2,2]*x[3,2] + 1,
	 x[1,3]*x[2,3]*x[3,3] + 1]
# Optimize semidefinite relaxation
val, model, _ = pcpop(0, 2; op_eq=R, lvl_lm=0)
println("Termination status ", termination_status(model))
\end{lstlisting}
\end{juliacode}

\subsection{Contextuality hypergraph}

A \emph{contextuality hypergraph} $H=(V, E)$ represents a measurement scenario with outcomes $V$ and measurements $E$. A quantum realization of $H$ is an assignment $P : V \to \mathcal P(H)$ of projectors in a Hilbert space $H$ that satisfies that $\sum_{v \in e} P(v) = \id$ for each hyperedge $e \in E$ \cite{acin2015combinatorial}. That is, quantum realizations are solutions of a polynomial optimization problem
\begin{align}
    \sup \quad & 0
\label{eq:hypergraph} \\
    \operatorname{s.t.} \quad & P_{v}P_{v} = P_v & v \in V \, , \nonumber \\
    & \sum_{v\in e} P_v = \id & e \in E \, . \nonumber
\end{align}
For instance, the contextuality hypergraph with $16$ vertices and $12$ edges in \cite[Figure 7]{acin2015combinatorial} corresponds with a bipartite Bell scenario with two dichotomic measurements. Therefore, the maximal quantum value for the contextuality inequality corresponding with the CHSH functional is $2\sqrt 2$. Namely,
\begin{align}
        \sup \quad & \sum_{a+b=xy} P_{ab|xy}
\label{eq:hypergraph_chsh} \\
    \operatorname{s.t.} \quad & P_{ab|xy}P_{ab|xy} = P_{ab|xy} \, , \nonumber \\
    & P_{00|xy} +  P_{01|xy} +  P_{10|xy} +  P_{11|xy} = \id \, , \nonumber \\
    & P_{00|x0} +  P_{01|x0} +  P_{10|x1} +  P_{11|x1} = \id \, , \nonumber \\
    & P_{00|x1} +  P_{01|x1} +  P_{10|x0} +  P_{11|x0} = \id \, , \nonumber \\
    & P_{00|0y} +  P_{10|0y} +  P_{01|1y} +  P_{11|1y} = \id \, , \nonumber \\
    & P_{00|1y} +  P_{10|1y} +  P_{01|0y} +  P_{11|0y} = \id \, . \nonumber
\end{align}
This value can be attained with the semidefinite relaxation of the corresponding polynomial optimization problem, where level 1 for the localizing matrix is used. The implementation in \texttt{PCPOP} is shown below.
\begin{juliacode}
\begin{lstlisting}[style=code, mathescape=true]
# Build the monoid
@pcmonoid M a[16,0]
Projector.(a)
A = reshape(a, 4, 4)
for i in 1:4
    @comms A[i, :]
    @comms A[:, i]
end
@comms union(A[1:2,1:2])
@comms union(A[1:2,3:4])
@comms union(A[3:4,1:2])
@comms union(A[3:4,3:4])
build(M)
# Objective function
p = A[2,2] + A[1,3] + A[3,1] + A[1,1]
p+= A[2,4] + A[4,2] + A[4,3] + A[3,4]
p+=-A[1,2] - A[1,4] - A[2,1] - A[2,3]
p+=-A[3,2] - A[3,3] - A[4,1] - A[4,4]
# Constraints
R = []
for i in 1:4
    append!(R, [one(M) - sum(A[i,:])])
    append!(R, [one(M) - sum(A[:,i])])
end
append!(R, [one(M) - sum(A[1:2,1:2])])
append!(R, [one(M) - sum(A[1:2,3:4])])
append!(R, [one(M) - sum(A[3:4,1:2])])
append!(R, [one(M) - sum(A[3:4,3:4])])
# Semidefinite relaxation
val, model, _, _ = pcpop(p, 1; op_eq = R)
println("Optimal value is   ", val)
\end{lstlisting}
\end{juliacode}
\subsection{Cycle contextuality scenarios}
\label{n-cycle-contextuality}

The $n$-cycle contextuality scenario consists of $n$ parties distributed among an $n$-cycle, such that operators acting on adjacent parties commute. Correlations in these scenarios when each party has one dichotomic observable has been analysed in \cite{n-contextuality}. In particular, the $4$-cycle recovers CHSH scenario and the $5$-cycle recovers KCBS scenario. The maximal quantum value of Klyachko-Can-Binicioglu-Shumovsky (KCBS) inequality~\cite{klyachko2008simple} corresponds with the optimal value of the polynomial optimization problem
\begin{align}
\label{eq:n-cycle-contextuality-simple}
        \inf \quad & x_0x_1 + x_1x_2 + x_2 x_3 + x_3 x_4 + x_4x_0 \\
    \operatorname{s.t.} \quad & x_i^2 = \id \, , \nonumber \\
    & x_i x_{i+1} = x_{i+1} x_{i} \, . \nonumber
\end{align}
The second level semidefinite relaxation gives the value $-3.9443$, which coincides with the quantum bound in \cite[Theorem 7]{n-contextuality} up to numerical precision. The implementation in \texttt{PCPOP} is shown below.
\begin{juliacode}
\begin{lstlisting}[style=code, mathescape=true]
# Build monoid
@pcmonoid M x[5,0]
Unipotent.(M.vertices)
for i in 1:5
    @comms x[i] x[(i%5)+1]
end
build(M)
# Optimize semidefinite relaxation
obj= sum(x[i]*x[(i%5)+1] for i in 1:5)
val,_ = pcpop(obj, 2; min=true)
println("Optimal value is   ", val)
\end{lstlisting}
\end{juliacode}

Now consider the $n$-cycle scenario where each party has two dichotomic observables $x_{i,0}$ and $x_{i, 1}$ that uses to play CHSH games with adjacents parties. The optimal quantum value of the joint $n$-cyclic CHSH game corresponds with the optimal value of the polynomial optimization problem
\begin{align}
\label{eq:n-cycle-contextuality}
        \sup \quad & \sum_i x_{i,0} x_{i+1, 0} + x_{i,0} x_{i+1, 1} + x_{i,1} x_{i+1, 0} - x_{i,1} x_{i+1, 1}\\
    \operatorname{s.t.} \quad & x_{i,j}^2 = \id \, , \nonumber \\
    & x_{i,j} x_{i+1,j} = x_{i+1,j} x_{i,j} \, . \nonumber
\end{align}
The second level relaxation has value $12.6026$, already manifesting that all CHSH games can not simultaneously attain the maximal quantum value. The implementation in \texttt{PCPOP} is shown below. We use this example in Chapter~\ref{ch:benchmarking} to benchmark the performance of \texttt{PCPOP} against other polynomial optimization packages (implemented with projectors instead of unipotents for the sake of the comparison). We remark that the equality constraints for odd-cycles scenarios do not admit finite Gr\"obner bases for any monomial ordering, while \texttt{PCPOP} implements alternative canonical forms for all the constraints involved in this scenarios. Therefore it comes with no surprise that \texttt{PCPOP} outperforms other non-specialized implementations based on general replacement rules.

\begin{juliacode}
\begin{lstlisting}[style=code, mathescape=true]
# Build monoid
n = 5
@pcmonoid M x[2*n,0]
Projector.(M.vertices)
x = reshape(x, n, 2)
for i in 1:n
    @comms x[i,:] x[(i%5)+1,:]
end
build(M)
# Optimize semidefinite relaxation
obj = sum([let (a, b) = (x[i,:], x[(i%n)+1,:]); (1-2*a[1])*(1-2*b[1]) + (1-2*a[1])*(1-2*b[2]) + (1-2*a[2])*(1-2*b[1]) - (1-2*a[2])*(1-2*b[2]) end for i in 1:n])
val,model,_ = pcpop(obj, 2; primal=true)
println("Optimal value is   ", val)
\end{lstlisting}
\end{juliacode}

\section{Conditional entropies}
\label{sec:conditional entropies}

The conditional quantum entropy $H(A|B)_\rho = S(\rho_{AB}) - S(\rho_{B})$ quantifies the amount of information needed to describe a quantum state $\rho_{AB}$ from its marginal $\rho_{B}$, where $S(\rho) = \tr \rho \log \rho$ is the von Neumann entropy. Conditional quantum entropies encode the security of different quantum cryptographic protocols. For instance, the asymptotic rate of randomness extraction \cite{miller2014universal} or quantum key distribution \cite{devetak2005distillation}. Polynomial optimization provides device independent bounds for the conditional quantum entropy \cite{brown2024device}. Consider a bipartite Bell scenario. The asymptotic rate of randomness that can be extracted from Alice's outcome $a$ on the fixed setting $x=0$ is lower bounded with \cite[Lemma 2.3]{brown2024device}
\begin{equation}
H(A|x=0,E) \geq \sum_{i=0}^{m-1} \frac{w_i}{t_i \ln 2} (1+V_i) \, .
\end{equation}
Here, $t_i$ and $w_i$ are the nodes and weights of the Gauss-Radau quadrature over $[0,1]$ with $m$ points and fixed end $t=1$; and $V_i$ is the optimal value of the polynomial optimization problem below in the hermitian variables $A_{a|x}$ and $B_{b|y}$ corresponding to the operators in the Bell experiment, plus non-hermitian variables $Z_{a}$. Namely,
\begin{align}
  V_i = \inf \quad & \sum_a A_{a|0} (Z_a + Z_a^* + (1 - t_i)Z_a^*Z_a + t_i Z_a Za^*)
\label{eq:bff_example} \\
    \operatorname{s.t.} \quad & A_{a|x}^2 = A_{a|x} \, , \nonumber \\
    & B_{b|y}^2 = B_{b|y} \, , \nonumber \\
    & [A_{a|x}, B_{b|y}] = 0 \, , \nonumber \\
    & [A_{a|x}, Z_a] = 0 \, , \nonumber \\
    & [B_{b|y}, Z_a] = 0 \, , \nonumber \\
    & Z_a Z_a^* \leq \alpha_i \, , \nonumber \\
    & Z_a^* Z_a \leq \alpha_i \, , \nonumber \\
    & p(ab|xy) = \rho (A_{a|x} B_{b|y}) \, . \nonumber \end{align}
Notice that the success of the protocol relies on some observed condition on the correlations $p(ab|xy)$, which enters the polynomial optimization as linear constraints over the moments. In this case, we assume that the correlations attain the maximal quantum value of CHSH functional. The lower bound on the conditional quantum entropy obtained with $m=8$ nodes and the second order semidefinite relaxation for Problems~\eqref{eq:bff_example} is $0.9887$. This certifies asymptotically at least $0.9887$ bits of randomness in Alice's outcome $a$ for the measurement setting $0$.
The implementation in \texttt{PCPOP} is shown below.
\begin{juliacode}
\begin{lstlisting}[style=code, mathescape=true]
using JuMP,Mosek,MosekTools,FastGaussQuadrature
function gaussradau(m)
    x, v = FastGaussQuadrature.gaussradau(m);
    t = 0.5*(1 .- x);
    w = 0.5*v;
    return (t, w)
end

k=2
m=8
t, w = gaussradau(m)
t = t[2:end]
w = w[2:end]
f(A,z,t) = A * (z + conj(z) + (1-t)*conj(z)*z) + t*z*conj(z)
γ = 2*(sqrt(2))
# Build monoid
@pcmonoid M Z[0, 2] a0 a1 b0 b1
z = M.vertices[1:2]
@comms [a0, a1] [b0, b1] z
Projector.([a0, a1, b0, b1])
build(M)
Id = one(M)
# Constraints CHSH violation B(p) ≥ γ
B  = (1-2*a0)*(1-2*b0) + (1-2*a0)*(1-2*b1)
B += (1-2*a1)*(1-2*b0) - (1-2*a1)*(1-2*b1)
tr_ge = [[B, γ]]
basis_principal = mons_at_level(M, k)
basis = basis_principal
H = 0.0
obj = f(a0, z[1], t[1]) + f(1-a0, z[2], t[1])

model, S, V, mons, LMI = npa_dual(obj, basis, basis_principal; 
    tr_ge=tr_ge, min=true, change_objective=true)

set_optimizer(model, Mosek.Optimizer)
optimize!(model)
ov1 = objective_value(model)
H += w[1]/(t[1]*log(2))*(1 + ov1)
old_obj = obj
for i in 2:length(t)
    obj = f(a0, z[1], t[i]) + f(1-a0, z[2], t[i])
    S = S + old_obj - obj
    model,V = model_new_obj(model, S, V, mons, LMI, -1)
    # set_silent(model)
    optimize!(model)
    ovi = objective_value(model)
    old_obj = obj
    H += w[i]/(t[i]*log(2))*(1 + ovi)
end
println("Conditional entropy lower bound : ", H)
\end{lstlisting}
\end{juliacode}

\section{Quantum networks}
\label{sec:networks}

One paradigmatic example of a quantum network that has been widely studied in the literature is the \emph{bilocal scenario} \cite{wolfe2019inflation, wolfe2021quantum, tavakoli2022bell, smith2026fully, renou2026two}. This scenario considers three parties that share two sources $\rho_{AB}$ and $\rho_{B'C}$, and perform measurements $A_{a|x}$, $B_{b|y}$ and $C_{c|z}$ over systems $A$, $BB'$ and $C$ respectively. We consider the following state polynomial optimization problem, corresponding to the maximal quantum value of Mermin inequality in the bilocal scenario with two measurement settings and two outcomes per party:
\begin{align}
\sup \ \ &   \rho(a_0b_0c_1 + a_0b_1c_0 + a_1b_0c_0 - a_1b_1c_1) \label{eq:bilocal}  \\
s.t. \ \ 
& a_i^2 = \id \, , \hspace{2em} [b_j, c_k] = 0 \, , \nonumber \\
& b_j^2 = \id \, , \hspace{2em} [a_i, c_k] = 0 \, , \nonumber \\
& c_k^2 = \id \, , \hspace{2em} [a_i, b_j] = 0 \, , \nonumber \\
& \rho(u(a_0, a_1) v(c_0, c_1)) = \rho(u(a_0, a_1))\rho(v(c_0, c_1)) \, . \nonumber
\end{align}
Here $u(a_0, a_1)$ and $v(c_0, c_1)$ run over all words in the letters $(a_0, a_1)$ and $(c_0, c_1)$ respectively. Similar constraints to capture the separability of states in causal networks have been proposed in \cite{pozas2019bounding, ligthart2023inflation, klep2024state, renou2026two}. The optimal value is $2\sqrt{2}$, which is attained with the second level semidefinite relaxation.
The implementation in \texttt{PCPOP} is shown below.
\begin{juliacode}
\begin{lstlisting}[style=code, mathescape=true]
using JuMP
# Build the monoid
@pcmonoid M a[2,0] b[2,0] c[2,0]
Unipotent.(M.vertices)
@comms a b c
build(M)
k = 2
TM = make_trace_monoid(M, 2*k, tracial=false) 
# Objective function.
p = state(a[1]*b[1]*c[2] + a[1]*b[2]*c[1], TM)
p+= state(a[2]*b[1]*c[1] - a[2]*b[2]*c[2], TM)
# Equality constraints
basis = trace_monomials(TM, 0:k)
wα = mons_at_level(a, k)
wγ = mons_at_level(c, k)
R = [state(u*v, TM) - state(u, TM)*state(v, TM) for u in wα for v in wγ]
R = unique([r for r in R if !(r==0)])
model = tpop(p, TM, basis, equalities=R)
set_optimizer(model, Mosek.Optimizer)
optimize!(model)
println("Termination status ", termination_status(model))
println("Optimal value is   ", val)
\end{lstlisting}
\end{juliacode}

\section{Uncertainty relations}
\label{sec:uncertainty}

State polynomial optimization can be used to characterize algebraic uncertainty relations \cite{moran2024uncertainty}. Consider, for example, the problem of finding the maximum of the sum of the squared expectation values of three unitary anti-commuting operators. In the state polynomial algebra over the variables $(x, y, z)$ with state symbol $\rho$, this problem reads
\begin{align}
\sup \ \ &  \rho(x)^2  + \rho(y)^2 + \rho(z)^2 \label{eq:ubs_paulis}  \\
s.t. \ \ & x^2 = \id\,, \hspace{1cm} yz = -zy \,, \nonumber \\
         & y^2 = \id\,, \hspace{1cm} zx = -xz \,, \nonumber \\
         & z^2 = \id\,, \hspace{1cm} xy = -yx \,. \nonumber
\end{align}
The semidefinite relaxation over the four dimensional subspace spanned by the state monomials $\{\id, x \rho (x), y \rho (y), z \rho(z)\}$ becomes
\begin{align}
1 = \sup \ \ & a + b + c 
\label{eq:ubs_lovasz_pauli} \\
\st & \begin{pmatrix}
1 & a & b & c \\
a & a & 0 & 0 \\
b & 0 & b & 0 \\
c & 0 & 0 & c
\end{pmatrix} \geq 0 \, . \nonumber
\end{align}
This is precisely the Lov\'asz number of the triangle graph and has optimal value $1$, which already matches the lower bound obtained with Pauli matrices. The implementation in \texttt{PCPOP} is shown below.
\begin{juliacode}
\begin{lstlisting}[style=code, mathescape=true]
using JuMP, Mosek, MosekTools
# Build base monoid in variables x, y, z
@pcmonoid M x y z
Unipotent.([x, y, z])
build(M)
# Build state monoid over M
TM = make_trace_monoid(M, 6, tracial=false)
# Objective function
ρx = state(x, TM)
ρy = state(y, TM)
ρz = state(z, TM)
p = ρx^2 + ρy^2 + ρz^2
# Anti-commutation relations
μx, μy, μz = TM.vertices_free
R = [μx*μy + μy*μx,
	 μy*μz + μz*μy,
	 μz*μx + μx*μz]
# Optimize semidefinite relaxation
basis = union(trace_monomials(TM, 0:1), [μx*ρx, μy*ρy, μz*ρz])
sos_model = tpop(p, TM, basis, equalities=R)
set_optimizer(sos_model, Mosek.Optimizer)
optimize!(sos_model)
println("Termination status ", termination_status(sos_model))
println("Optimal value is   ", objective_value(sos_model))
\end{lstlisting}
\end{juliacode}

\section{Almost qudits}
\label{sec:dimension}

The framework proposed in \cite{pauwels2022almost} allows to quantify the effects of the assumptions on the physical dimension in certain quantum information protocols. An \emph{almost qudit} is a state whose support is almost contained in a $d$-dimensional space. 
Correlations in prepare and measurement scenarios with almost qudits can be approximated with semidefinite programs. Let $\rho_{x_1 x_2}$ be an almost qubit and $M_{b|y}$ projective measurement effect where each $x_1, x_2, b, y$ is a bit. The randomness in $b$ for a fixed setting $x_1,x_2,y=1$ conditioned to a random access code value $ \sum p(x_y|x_1x_2y) = c$ is given by the guessing probability, which can be approximated with a tracial polynomial optimization problem, with normalization $\tau(\rho_{x_1x_2}) = 1$ instead of $\tau(\id) = 1$ ($\tau$ is the tracial state symbol). Namely,
\begin{align}
    \label{eq:almost_qubits} \mathds P_g(b) \, = \, \sup \, & \rho_{11} M_{b|1} \\
    \operatorname{s.t.} \, 
    & \rho_{x_1 x_2}^2 = \rho_{x_1 x_2} \, , \nonumber \\
    & M_{b|y}^2 = M_{b|y} \, , \nonumber \\
    & \Pi^2 = \Pi \, , \nonumber \\
    &\tau(\rho_{x_1 x_2}) = 1 \, , \nonumber\\
    &\tau(\Pi) = d \, , \nonumber\\
    &\tau(\rho_{x_1x_2} \Pi) \geq 1 - \epsilon \, , \nonumber \\
    & \textstyle \sum \tau(\rho_{x_1x_2} M_{x_y|y}) = c \, . \nonumber
\end{align}
The third level relaxation for Problem~\eqref{eq:almost_qubits} conditioned to the maximal random access code value $c = 2(2+\sqrt{2})$ with $d=2$ and $\epsilon = 0.01$ has optimal value $0.9733$. Therefore, the randomness certified in the protocol significantly decreases from $0.228$ bits with perfect qubits to $0.039$ bits with $\epsilon$-approximate qubits [see \cite[Figure 1]{pauwels2022almost}]. The implementation in \texttt{PCPOP} is shown bellow.
\begin{juliacode}
\begin{lstlisting}[style=code, mathescape=true]
#Parameters
d = 2
ϵ = 0.01
c = 2*(2+sqrt(2))

# Build monoid
@pcmonoid M ρ[4, 0] B[2, 0] P[1, 0]
Projector.(ρ)
Projector.(B)
Projector.(P)
build(M)

# Objective function
obj = ρ[1]*B[1]
# Linear equalities on the moments
rac = ρ[1]*B[1] + ρ[2]*B[1] + ρ[3]*(1-B[1]) + ρ[4]*(1-B[1])
rac+= ρ[1]*B[2] + ρ[2]*(1-B[2]) + ρ[3]*B[2] + ρ[4]*(1-B[2])
tr_eq = [[ρ[1], 1], [ρ[2], 1], [ρ[3], 1], 
         [ρ[4], 1], [P[1], d], [rac, c]]
# Linear inequalities on the moments
tr_ge = [[ρ[1]*P[1], 1 - ϵ],
         [ρ[2]*P[1], 1 - ϵ], 
         [ρ[3]*P[1], 1 - ϵ], 
         [ρ[4]*P[1], 1 - ϵ]]

# Optimization of the semidefinite relaxation
val, model, _ = pcpop(obj, 3; tr_eq = tr_eq,
    tr_ge = tr_ge,
    tracial = true,
    normalize = false,
)
println("Optimal value      ", val)

\end{lstlisting}
\end{juliacode}

\section{Information capacity}
\label{sec:capacity}

We consider the prepare and measure scenario with constraints on the communication proposed in \cite{tavakoli2022informationally}. In the prepare and measure scenario $(X, Y, B)$, for each input value $x \in X$, the sender prepares a physical state $\rho_x$ that sends to the receiver, who performs a measure chosen with the input value $y \in Y$ and obtains an outcome $b\in B$. A constraint on the communication appears as an upper bound on the probability to guess the input value $x$, which is simply the maximal discrimination probability for the states $\rho_x$ when $x$ are uniformly distributed. Namely, $P_g(X) \leq G$ for some $G \in [0,1]$.

Classical correlations are the feasible solutions of a linear program, while quantum correlations are the feasible solutions of a tracial polynomial optimization problem, which can be approximated with semidefinite programs. 
We consider the bounds obtained in \cite[\S 4.3]{tavakoli2022informationally}, which correspond with the problem \cite[Equation 76]{tavakoli2022informationally} for the scenario $(3,2,2)$ with the linear witness in \cite[Equation 46]{tavakoli2022informationally} and uniformly distributed $x$.

\begin{align}
    \sup \, & - \tau(\rho_0 a_{0}) - \tau(\rho_0 a_{1}) - \tau(\rho_1 a_0) + \tau(\rho_1a_1) + \tau(\rho_2a_0) \\
    \operatorname{s.t.} \, 
    & \tau(\rho_x) = 1 \, , \nonumber \\
    & \tau(\sigma) \leq G \, , \nonumber \\
    & a_i^2 = \id \, , \nonumber \\
    & \rho_x \geq \rho_x^2 \, , \nonumber \\
    & \sigma \geq \rho_x/3 \, . \nonumber
\end{align}
Here, $\rho_x$ denotes the state prepared by the sender for the input $x$, $a_i$ the dichotomic measurements performed by the receiver, and $\tau$ the tracial state with normalization $\tau(\rho_x) = 1$ instead of $\tau(\id) = 1$. The auxiliary operator $\sigma$ incorporates the constraints on the communication.
The optimal value for this expression over classical correlations is $6G-1$, which is $3.8$ for $G=0.8$. The second level semidefinite relaxation for the quantum correlations with $G=0.8$ has value $4.4128$. The implementation in \texttt{PCPOP} is shown below.

\begin{juliacode}
\begin{lstlisting}[style=code, mathescape=true]
G = 0.8
# Build the monoid
@pcmonoid M ρ[3,0] σ a[2,0]
Unipotent.(a)
build(M)
# Conditions on the operators
op_ge = vcat([σ-(1 /3)*r for r in ρ], [r-r^2 for r in ρ]) 
# Conditions on the moments
tr_ge = [[-σ,-G]]
tr_eq = [[ρ[x],1] for x in 1:3]
# Optimization of the semidefinite relaxation
obj = -a[1]*ρ[1] - a[2]*ρ[1] - a[1]*ρ[2] +  a[2]*ρ[2] + a[1]*ρ[3]
val, model, _ = pcpop(obj, 2; op_ge=op_ge,
                 tr_eq=tr_eq,
                 tr_ge=tr_ge,
                 tracial=true,
                 normalize=false)
println("Optimal value is ", val)
\end{lstlisting}
\end{juliacode}
\chapter{Benchmarking}
\label{ch:benchmarking}

In this Chapter we benchmark \texttt{PCPOP} with a variety of problems and compare the performance with other state of the art packages. We divide these problems into two sections:
\begin{enumerate}[noitemsep]
    \item \textbf{Algebraic computations}. These problems include computing canonical forms of words with respect to given constraints and performing arithmetical computations with them, such as multiplication. These algebraic computations are expected to be performed a large number of times in any practical application, so it is essential to guarantee a competitive performance. 
    \item \textbf{Polynomial optimization}.
    Natural benchmarking parameters for polynomial optimization purposes include the cost of building the semidefinite programming relaxations of a given problem and the size of the resulting relaxations, which automatically reflects in the cost of solving the semidefinite program.
\end{enumerate}

All the results displayed in this chapter are obtained with 20 cores 64GB memory 12th Gen Intel(R) Core(TM) i7-12700.

\section{Algebraic computations}

We compare the performance of \texttt{PCPOP}, \texttt{OSCAR} and \texttt{QuantumNPA} doing arithmetical computations and finding canonical forms in different scenarios.
\texttt{OSCAR}~\cite{OSCAR} is a sophisticated general-purpose computer algebra system implemented in Julia, which can compute Gr\"obner bases and implement subsequent reductions in different algebras among many other functionalities. In order to obtain canonical forms with respect to a given collection of constraints, \texttt{OSCAR} first computes a Gr\"obner basis which is later used to perform reductions.
\texttt{QuantumNPA}~\cite{quantumnpa} is a specific implementation in Julia of the semidefinite programming relaxations for non-commutative polynomial optimization problems. It provides effective canonical forms for some physically inspired constraints, such as commutations, projections, unitaries and unipotents.

\begin{figure}[t!]
    \centering
    \includegraphics{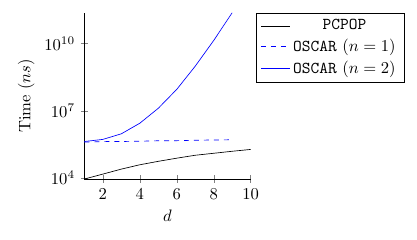}
	\caption{\textbf{Benchmarking building time} of the partially commutative monoid in \texttt{PCPOP} (black) and the Gr\"obner basis in \texttt{OSCAR} (blue) truncated to degree $d$ for the scenarios with $n$ parties in Example~\eqref{ex:partial_comms}.}
    \label{fig:benchmark_canonical_form}
\end{figure}

\begin{figure*}[t!]
    \centering
    \includegraphics{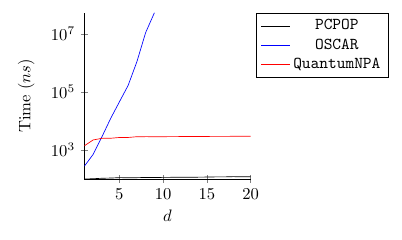}
	\caption{\textbf{Benchmarking computing canonical forms} with respect to the commutation relations in Example~\ref{ex:partial_comms} in \texttt{PCPOP} (black), \texttt{OSCAR} (blue) and \texttt{QuantumNPA} (red). The figure shows the time in nanoseconds of computing the canonical form of $wv$ for random words $w$ and $v$ of length $d$ in $3n=6$ variables $(\vec a, \vec b, \vec c)$.}
    \label{fig:partial_comms_reduction}
\end{figure*}

\begin{figure*}[t!]
    \centering
    \includegraphics{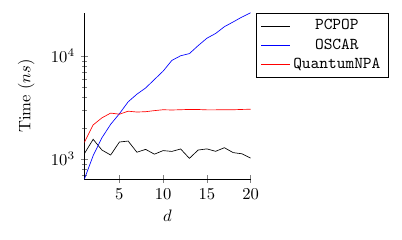}
    \caption{\textbf{Benchmarking computing canonical forms} with respect to the commutation relations and projection constraints in Example~\ref{ex:partial_comms} in \texttt{PCPOP} (black), \texttt{OSCAR} (blue) and\texttt{QuantumNPA} (red). The figure shows the time in nanoseconds of computing the canonical form of $wv$ for random words $w$ and $v$ of length $d$ in $3n=6$ variables $(\vec a, \vec b, \vec c)$.}
        \label{fig:partial_comms_projs_reduction}
\end{figure*}

\begin{example}
	\label{ex:partial_comms}
	Let $\vec a$, $\vec b$ and $\vec c$ be collections of $n$ variables each, such that variables in $\vec a$ commute with variables in $\vec b$ and variables in $\vec b$ commute with variables in $\vec c$, but variables in $\vec a$ do not commute with variables in $\vec c$. We compare in Figure~\ref{fig:benchmark_canonical_form} the cost of computing a level $d$ truncated Gr\"obner basis in \texttt{OSCAR} and building the partially commutative monoid in \texttt{PCPOP} for both scenarios. Subsequently, we consider the problem of computing canonical forms with respect to the commutation relations. 
    We compare in Figure~\ref{fig:partial_comms_reduction} the average cost of performing the multiplication $wv$ of two random words $w$ and $v$ of length $d$ and later obtaining the canonical form. Figure~\ref{fig:partial_comms_projs_reduction} shows the same comparison in the scenario with the additional constraints that each variable is a projector.
\end{example}

The results show that the implementation in \texttt{PCPOP} specialized for partially commutative computations significantly outperforms the general-purpose implementation in \texttt{OSCAR} through Gr\"obner basis reductions.

\section{Polynomial optimization}

We benchmark the performance of \texttt{PCPOP} against the polynomial optimization packages \texttt{Ncpol2sdpa} (Python), \texttt{QuantumNPA} (Julia) and \texttt{Moment} (C++ with Matlab interface). In particular, we compare the size of the semidefinite relaxations, the number of variables and constraints, the set up time, the solving time and the optimal value for three selected problems: the maximal quantum value of CHSH functional [Equation~\ref{eq:chsh}], bounds for the conditional quantum entropy [Equation~\eqref{eq:bff_example}] and the $n$-cycle contextuality problem [Equation~\eqref{eq:n-cycle-contextuality}]. The results of the benchmarking are presented in Tables~\ref{tab:benchmark_chsh}, \ref{tab:benchmark_bff}, \ref{tab:benchmark_polygon_bell} and \ref{tab:benchmark_polygon_bell2}. 

\begin{table}[h]
\begin{tabular}{llccccccc}
\centering
\renewcommand{\arraystretch}{1.1}
$d$ & Package &SDP size &\# cons &\# vars & setup (s) & solve (s) \\
\hline

\multirow{4}{*}{12}
& \texttt{Ncpol2sdpa} &\multirow{4}{*}{313} &\multirow{4}{*}{780} &\multirow{4}{*}{49141} &11451.0598 &0.8741 \\
& \texttt{QuantumNPA} & & & &0.7247 &0.8177 \\
& \texttt{Moment}     & & & &0.0778 &0.9179 \\
& \texttt{PCPOP}      & & & &0.5518 &0.9497 \\
\hline

\multirow{4}{*}{13}
& \texttt{Ncpol2sdpa} &\multirow{4}{*}{365} &\multirow{4}{*}{910} &\multirow{4}{*}{66795} & & \\
& \texttt{QuantumNPA} & & & &1.0263 &1.1456 \\
& \texttt{Moment}     & & & &0.0933 &1.0995 \\
& \texttt{PCPOP}      & & & &0.8090 &1.1523 \\
\hline

\multirow{4}{*}{14}
& \texttt{Ncpol2sdpa} &\multirow{4}{*}{421} &\multirow{4}{*}{1050} &\multirow{4}{*}{88831} & & \\
& \texttt{QuantumNPA} & & & &1.3742 &1.5807 \\
& \texttt{Moment}     & & & &0.1144 &1.3005 \\
& \texttt{PCPOP}      & & & &1.164 &1.568 \\
\hline

\multirow{4}{*}{15}
& \texttt{Ncpol2sdpa} &\multirow{4}{*}{481} &\multirow{4}{*}{1200} &\multirow{4}{*}{115921} & & \\
& \texttt{QuantumNPA} & & & &1.8133 &2.3776 \\
& \texttt{Moment}     & & & &0.1422 &1.6668 \\
& \texttt{PCPOP}      & & & &1.5679 &2.3189 \\
\hline

\multirow{4}{*}{16}
& \texttt{Ncpol2sdpa} &\multirow{4}{*}{545} &\multirow{4}{*}{1359} &\multirow{4}{*}{148785} & & \\
& \texttt{QuantumNPA} & & & &2.4562 &3.593 \\
& \texttt{Moment}     & & & &0.1882 &3.6665 \\
& \texttt{PCPOP}      & & & &2.0383 &3.4822 \\
\hline
\end{tabular}
\caption{\textbf{Benchmarking polynomial optimization} of problems in Equations~\eqref{eq:chsh} with \texttt{Ncpol2sdpa}, \texttt{QuantumNPA}, \texttt{Moment} and \texttt{PCPOP}. For a relaxation of level $d$ we use the SDP size, the number of variables, the number of constraints, the setup time and the solving time as benchmarking parameters. The SDP size, number of constraints and number of variables are the same for all the packages.}
\label{tab:benchmark_chsh}
\end{table}

\begin{table}[h]
\begin{tabular}{llccccccc}
\centering
\renewcommand{\arraystretch}{1.1}
$d$ & Package &
SDP size &
\# cons &
\# vars &
setup (s) &
solve (s) \\
\hline

\multirow{4}{*}{2}
& \texttt{Ncpol2sdpa} &\multirow{4}{*}{49} &498 &1275 &1.1743 &1.5344 \\
& \texttt{QuantumNPA} & &498 &1226 &\multicolumn{2}{c}{0.4165} \\
& \texttt{Moment}     & &499 &1225 &0.0087 &0.7209 \\
& \texttt{PCPOP}      & &499 &1225 &0.0149 &0.3016 \\
\hline

\multirow{4}{*}{3}
& \texttt{Ncpol2sdpa} &\multirow{4}{*}{221} &7728 &24753 &17.9911 &196.6395 \\
& \texttt{QuantumNPA} & &7728 &24532 &\multicolumn{2}{c}{112.0566} \\
& \texttt{Moment}     & &7729 &24531 &0.0965 &88.8165 \\
& \texttt{PCPOP}      & &7729 &24531 &0.2497 &94.7844 \\
\hline

\multirow{4}{*}{4}
& \texttt{Ncpol2sdpa} & \multicolumn{5}{c}{-} \\
& \texttt{QuantumNPA} & \multirow{3}{*}{925} &122016 &428276 &7.4610 &\multirow{3}{*}{-} \\
& \texttt{Moment}     & &122017 &428275 &1.8627 & \\
& \texttt{PCPOP}      & &122017 &428275 &4.4878 & \\
\hline
\end{tabular}
\caption{\textbf{Benchmarking polynomial optimization} for the bounds on the relative quantum entropy in Equation~\eqref{eq:bff_example} with \texttt{Ncpol2sdpa}, \texttt{QuantumNPA}, \texttt{Moment} and \texttt{PCPOP}. For a relaxation of level $d$ we use the SDP size, the number of variables, the number of constraints, the setup time and the solving time (these are merged together for \texttt{QuantumNPA} to provide a fair comparison for technical differences in the implementation) as benchmarking parameters. The SDP size is the same for all the packages.}
\label{tab:benchmark_bff}
\end{table}

\begin{table}[h]
\begin{tabular}{llccccccc}
\centering
\renewcommand{\arraystretch}{1.1}
$n$ & Package &
SDP size &
\# cons &
\# vars &
setup (s) &
solve (s) \\
\hline

\multirow{4}{*}{8}
& \texttt{Ncpol2sdpa} &\multirow{4}{*}{225} &18536 &\multirow{4}{*}{25425} &11.2094 &52.4779 \\
& \texttt{QuantumNPA} & &18480 & &0.5018 &53.2079 \\
& \texttt{Moment}     & &18481 & &0.4528 &67.0534 \\
& \texttt{PCPOP}      & &6945 & &0.3354 &7.1483 \\
\hline

\multirow{4}{*}{9}
& \texttt{Ncpol2sdpa} &\multirow{4}{*}{289} &31884 &\multirow{4}{*}{41905} &19.7301 &387.9730 \\
& \texttt{QuantumNPA} & &31824 & &0.8364 &380.6007 \\
& \texttt{Moment}     & &31825 & &0.7461 &495.4557 \\
& \texttt{PCPOP}      & &10081 & &0.5521 &19.6686 \\
\hline

\multirow{4}{*}{10}
& \texttt{Ncpol2sdpa} &\multirow{4}{*}{361} &51364 &\multirow{4}{*}{65341} &32.7658 &1001.8819 \\
& \texttt{QuantumNPA} & &51300 & &1.7051 &1207.2346 \\
& \texttt{Moment}     & &51301 & &1.4559 &1322.0114 \\
& \texttt{PCPOP}      & &14041 & &0.9182 &36.4585 \\
\hline

\multirow{4}{*}{11}
& \texttt{Ncpol2sdpa} &\multirow{4}{*}{441} &78608 &\multirow{4}{*}{97461} &49.5094 &\multirow{3}{*}{-} \\
& \texttt{QuantumNPA} & &78540 & &2.0935 & \\
& \texttt{Moment}     & &78541 & &2.4256 & \\
& \texttt{PCPOP}      & &18921 & &1.7324 &108.5885 \\
\hline
\end{tabular}
\caption{\textbf{Benchmarking polynomial optimization} for the $n$-cycle contextuality scenario in Equation~\eqref{eq:n-cycle-contextuality-simple} with \texttt{Ncpol2sdpa}, \texttt{QuantumNPA}, \texttt{Moment} and \texttt{PCPOP}. For a relaxation of level $d$ we use the SDP size, the number of variables, the number of constraints, the setup time and the solving time as benchmarking parameters. The SDP size and the number of variables are the same for all the packages.}
\label{tab:benchmark_polygon_bell}
\end{table}

\begin{table}[h]
\begin{tabular}{llccccccc}
\centering
\renewcommand{\arraystretch}{1.1}
$n$ &$d$ & Package &
SDP size &
\# cons &
\# vars &
setup (s) &
solve (s) \\
\hline

\multirow{4}{*}{4} & \multirow{4}{*}{3}
& \texttt{Ncpol2sdpa} &233 &9748 &27261 &14.7382 &11.42 \\
& & \texttt{QuantumNPA} &\multirow{3}{*}{217} &\multirow{3}{*}{6156} &\multirow{3}{*}{23653} &0.2569 &4.3481 \\
& & \texttt{Moment}     & & & &0.4517 &5.9332 \\
& & \texttt{PCPOP}      & & & &0.1856 &5.0222 \\
\hline

\multirow{4}{*}{4} & \multirow{4}{*}{4}
& \texttt{Ncpol2sdpa} &1073 &184914 &576201 &500.8337 &\multirow{4}{*}{-} \\
& & \texttt{QuantumNPA} &\multirow{3}{*}{865} &\multirow{3}{*}{71281} &\multirow{3}{*}{374545} &5.7093 & \\
& & \texttt{Moment}     & & & &3.0951 & \\
& & \texttt{PCPOP}      & & & &2.4075 & \\
\hline

\multirow{4}{*}{5} & \multirow{4}{*}{3}
& \texttt{Ncpol2sdpa} &555 &76961 &154290 &84.0874 &\multirow{4}{*}{-} \\
& & \texttt{QuantumNPA} &\multirow{3}{*}{531} &\multirow{3}{*}{66516} &\multirow{3}{*}{141246} &2.5564 & \\
& & \texttt{Moment}     & & & &3.9535 & \\
& & \texttt{PCPOP}      & & & &1.7831 & \\
\hline

\multirow{4}{*}{5} & \multirow{4}{*}{4} 
& \texttt{Ncpol2sdpa} & & & - & & \\
& & \texttt{QuantumNPA} &\multicolumn{5}{c}{-} \\
& & \texttt{Moment}     &\multicolumn{5}{c}{-} \\
& & \texttt{PCPOP}      &3361 &2562002 &5649842 & 93.3293 & - \\
\hline
\end{tabular}
\caption{\textbf{Benchmarking polynomial optimization} of the $n$-cycle contextuality problem in Equation~\eqref{eq:n-cycle-contextuality} with \texttt{Ncpol2sdpa}, \texttt{QuantumNPA}, \texttt{Moment} and \texttt{PCPOP}. In this scenario each party can perform two measurements, which we encode with two projectors. For a relaxation of level $d$ we use the SDP size, the number of variables, the number of constraints, the setup time and the solving time as benchmarking parameters. The SDP size, number of variables and number of constraints are the same for \texttt{QuantumNPA}, \texttt{Moment} and \texttt{PCPOP}.}
\label{tab:benchmark_polygon_bell2}
\end{table}

For the examples benchmarked in Tables~\ref{tab:benchmark_chsh} and~\ref{tab:benchmark_bff}, the setup times are the smallest with \texttt{Moment} and the largest with \texttt{Ncpol2sdpa}, although the size of the relaxations and setup times are similar for all packages except for \texttt{Ncpol2sdpa} and the solving times are similar. 
The example benchmarked in Table~\ref{tab:benchmark_polygon_bell} shows that the setup times for \texttt{PCPOP}, \texttt{QuantumNPA} and \texttt{Moment} are comparable, but \texttt{PCPOP} is faster in solving.
Notice that all constraints in this example are internally set in the monoid. Therefore, the semidefinite relaxation only involves one positive semidefinite matrix. The difference in the number of constraints is due to the implementation: \texttt{PCPOP} introduces linear constraints identifying all equivalent monomials in the moment matrix, while the other packages create one linear constraint for each equivalence class of monomials (this coincides with the sum of squares implementation in \texttt{PCPOP}).
The example benchmarked in Table~\ref{tab:benchmark_polygon_bell2} shows comparable building times for \texttt{PCPOP}, \texttt{QuantumNPA} and \texttt{Moment}, \texttt{PCPOP} being faster. In particular, only \texttt{PCPOP} succeeds in building the semidefinite relaxation in the last instance of the table. 
Moment implements equality constraints through substitution rules computed by a Knuth–Bendix completion procedure \cite{knuth1970simple}. In polynomial algebras, this completion process is closely related to Gröbner basis computation, with the resulting substitution rules corresponding to a truncated Gröbner basis. The size of the substitution system can be limited to a desired threshold $k$ using \texttt{setting.complete(k)}, which automatically reflects on the building times for \texttt{Moment}. We do not fix $k$ for Table~\ref{tab:benchmark_chsh}, we set $k = 20$ for Table~\ref{tab:benchmark_bff}. For Table~\ref{tab:benchmark_polygon_bell}, we set $k = 100$ for $n=8$  and we set $k = 150$ for other values of $n$. For Table~\ref{tab:benchmark_polygon_bell2}, we set $k=100$.

\chapter{Concluding words}
\label{ch:conclusion}

\texttt{PCPOP} is an open-source and multi-purpose package in Julia for polynomial optimization. The package automatically builds and solves semidefinite programming approximations to non-commutative, tracial, trace and state polynomial optimization problems. \texttt{PCPOP} additionally implements the recently developed specialized framework of partially commutative polynomial optimization, which relies on alternative representations for partially commutative monomials and is especially effective for problems appearing in quantum information. 
Moreover, the package offers several functionalities such as algebraic reductions, symmetry reductions and Jordan algebra reductions for the semidefinite approximations. Although further sparsity reductions can in principle be specialized to the partially commutative setting, these are not implemented in the current release.
In addition to a competitive implementation, \texttt{PCPOP} provides an user-friendly interface with flexible methods that cover a variety of problems. We have collected some examples that appear in state-of-the-art research in the field of quantum information. 
\FloatBarrier
\chapter{Acknowledgements}

We thank Stefano Pironio for his valuable feedback and suggestions to improve the package. We thank and Beno\^{\i}t Legat for advice with the implementation and telling us about the package \texttt{SymbolicWedderburn}. We thank Erik Woodhead for technical discussions on the implementation of the moment and sum of squares relaxations in \texttt{QuantumNPA}. We thank Mateus Ara\'ujo for practical discussions on the implementation in \texttt{Moment} and references for the benchmarking.

\bibliographystyle{abbrv}
\bibliography{references}

\end{document}